\documentclass[reprint,twocolumn,showpacs,nofootinbib,notitlepage,preprintnumbers,longbibliography]{revtex4-2}
\pdfoutput=1
\usepackage[T1]{fontenc}
\usepackage{amsfonts}
\usepackage{amsmath}
\usepackage{amssymb}
\usepackage{multirow}
\usepackage{graphicx}
\usepackage{xcolor}
\usepackage{mathtools}
\usepackage{comment}

\def\be{\begin{equation}}
\def\ee{\end{equation}}
\def\bea{\begin{eqnarray}}
\def\eea{\end{eqnarray}}

\newcommand{\Tr}{{\rm {Tr}}}

\let\OriginalIncludeGraphics\includegraphics
\renewcommand{\includegraphics}[2][]{%
  \IfFileExists{#2}{%
    \OriginalIncludeGraphics[#1]{#2}%
  }{%
    \fbox{\scriptsize Missing image: \detokenize{#2}}%
  }%
}

\newcommand{\eqimage}[2][1.0cm]{%
	\raisebox{-0.4\height}{%
		\includegraphics[height=#1]{images/#2}%
	}%
}

\newcommand{\colormatrices}[2][1.5cm]{%
	\raisebox{-0.4\height}{%
		\includegraphics[width=#1]{images/color_matrices/#2}%
	}%
}

\newcommand{\graphimage}[2]{%
	\raisebox{-0.5\height}{%
		\includegraphics[width=#1]{images/#2}%
	}%
}

\newcommand{\melon}{%
	\eqimage{bubbles/bubble_rank3_n1_001.png}%
}
\newcommand{\pillowone}{%
	\eqimage{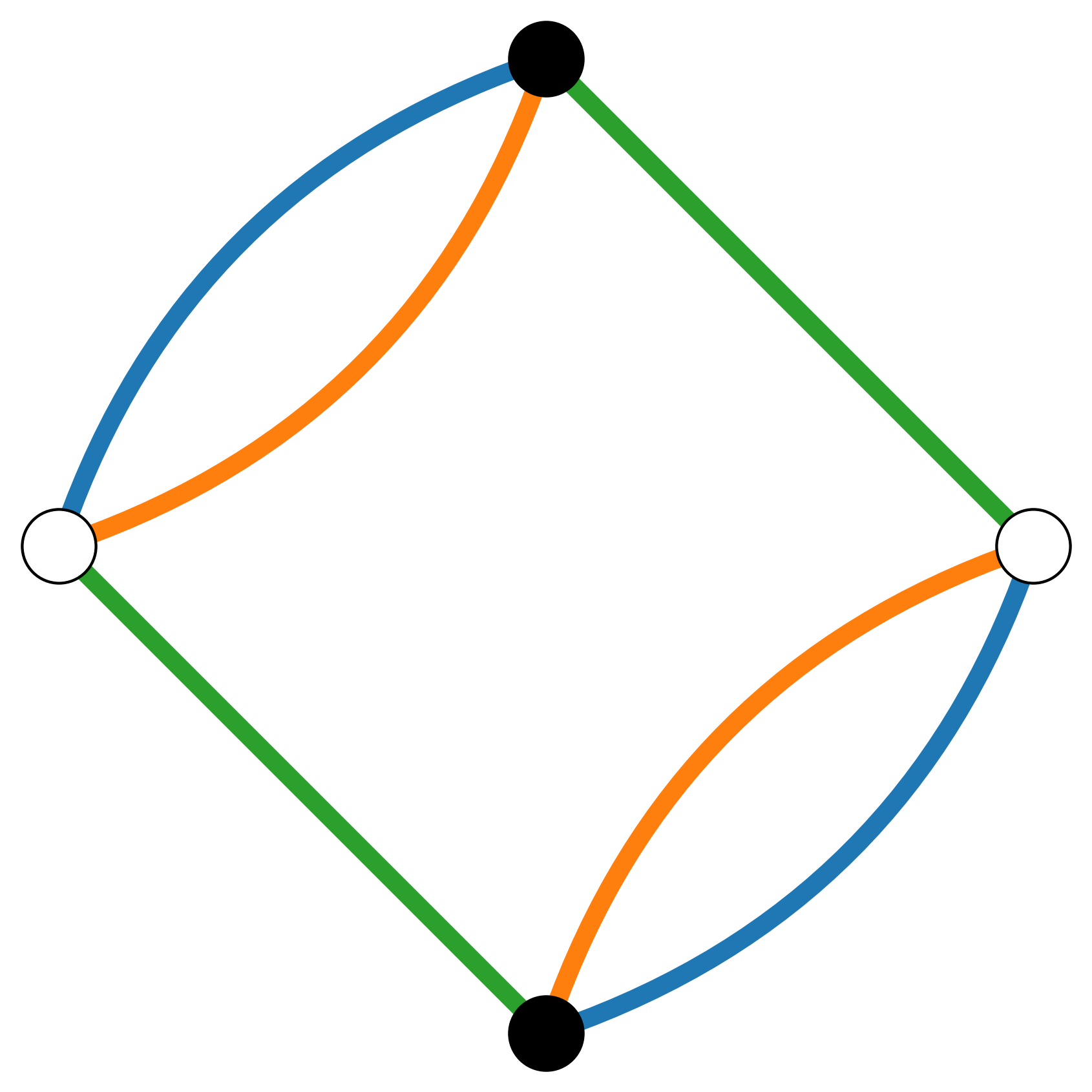}%
}
\newcommand{\pillowtwo}{%
	\eqimage{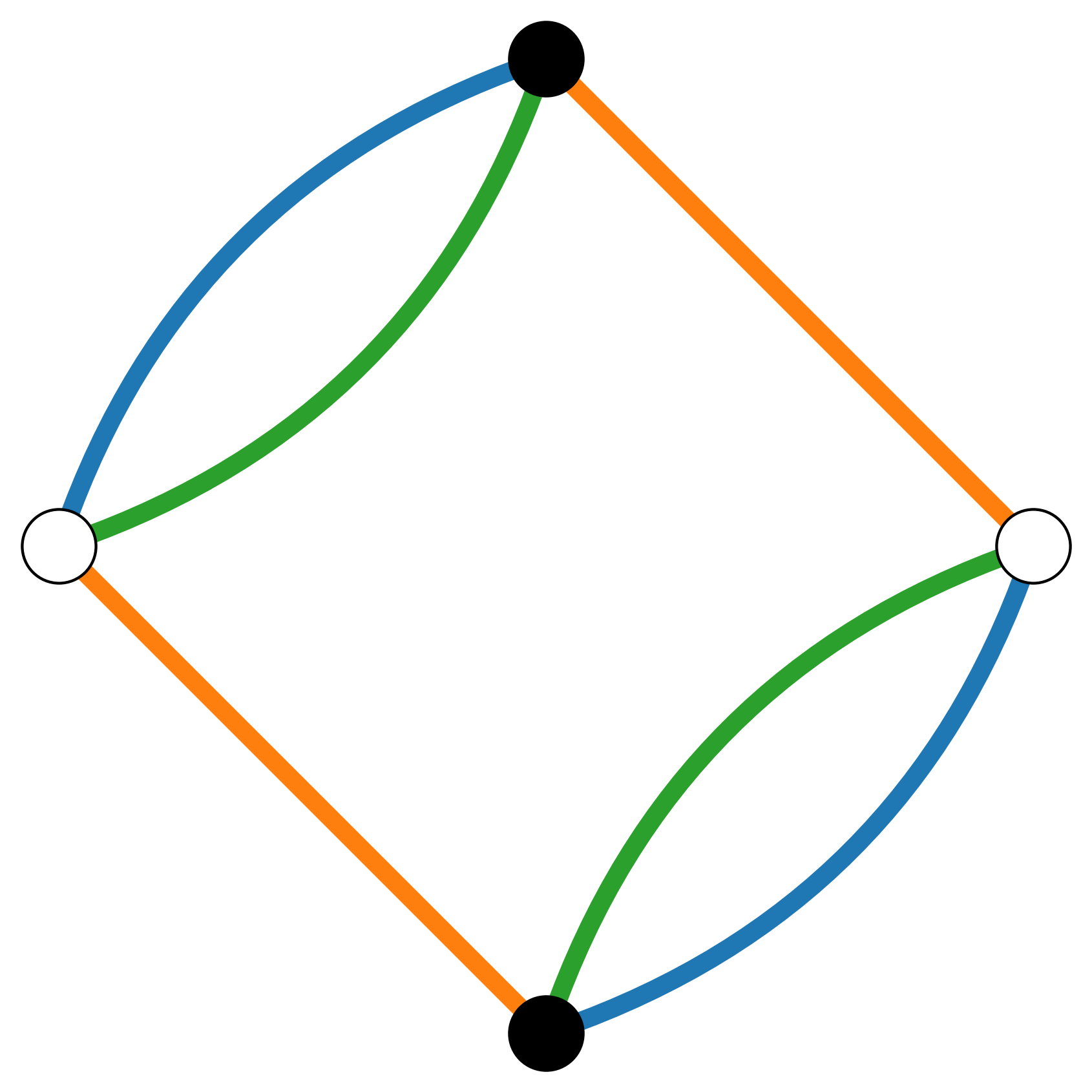}%
}
\newcommand{\pillowthree}{%
	\eqimage{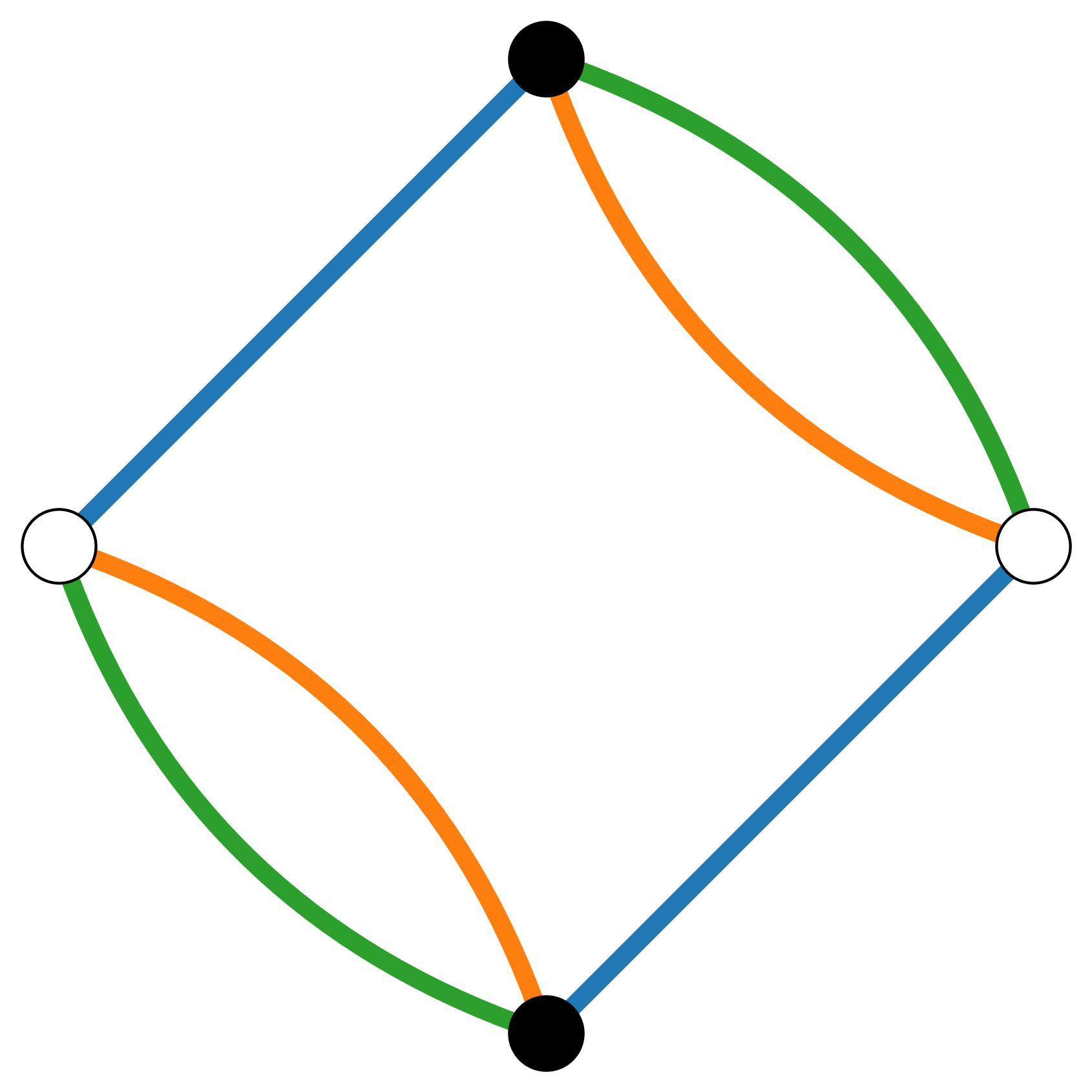}%
}

\begin{document}

\title{Additional constraints for the tensor bootstrap}

\author{Samuel Laliberte}
\email{samuel.laliberte@oist.jp}

\author{Reiko Toriumi}
\email{reiko.toriumi@oist.jp}

\affiliation{Okinawa Institute of Science and Technology Graduate University, 1919-1, Tancha, Onna, Kunigami District, Okinawa 904-0495, Japan}

\date{\today}

\begin{abstract}
Recently, new positivity constraints were suggested to constrain arbitrary unitary tensor integrals. In the present work, we explore two variants of these positivity constraints: one built from ``open bubbles'', which are tensor-like objects found by removing a tensor from a bubble invariant, and the second built from ``color matrices'', which are matrices found by removing a color contraction from a bubble invariant. Using these positivity constraints, we find sharp bounds on unitary tensor integrals at finite $N$, and probe deviations from Gaussian universality in this limit.
\end{abstract}

\maketitle

\section{Introduction}

Tensor models provide a natural framework to study interacting systems with many degrees of freedom organized into higher-rank tensors. Over the past decades, they have found several applications in physics. Notably, tensor models have been explored as candidates to generalize the success of matrix models in describing 2D quantum gravity to higher dimensions \cite{Sasakura:1990fs, Ambjorn:1990ge, Gross:1991hx, Gurau:2011xp, gurau2017random, 
Gurau:2013cbh, Carrozza:2015adg, Bonzom:2016kqf, Casali:2017tfh, 
Ryan:2011qm, Martini:2021rbr, Lionni:2017xvn, chandra2024stochasticanalysisapproachtensor}. They have also been used to define a version of the SYK model without disorder \cite{Sachdev:1992fk, Kitaev:2015talk, Maldacena:2016hyu, Witten:2016iux, Klebanov:2016xxf,Klebanov:2018fzb, 
Bonzom:2018jfo, Gurau_2017,  Bonzom_2017, delporte2018tensortrackvholographic, Ben_Geloun_2018}, where applications to quantum gravity in this context have been extensively studied. The various simplifications related to tensor models in the large $N$ limit have also proven useful to study field theories, their renormalizability, and conformal field theories (CFTs) \cite{BenGeloun:2011rc, Carrozza:2012uv, 
Ousmane_Samary_2015, Benedetti_2015, Krajewski:2016svb, Eichhorn:2017xhy, BenGeloun:2017xbd, Benedetti:2020nI, Benedetti_2020, Benedetti_2022, Jepsen:2023pzm}. More recently, tensor models have also found an application in the study of ensemble averages of CFTs \cite{Maloney:2020nni, Belin:2020hea, Anous:2021caj, Chandra:2022bqq, Sasieta:2022ksu, Belin:2023efa, deBoer:2024mqg, Jafferis:2025vyp, deBoer:2025rct, Jafferis:2025yxt, Hartman:2025ula, Chandra:2025fef, Wang:2025jgo, Belin:2026pko}, where they can be used to average over OPE coefficients \cite{Belin:2023efa, Jafferis:2025vyp, Jafferis:2025yxt}. It was also found that tensors provide a novel way to classify multipartite entanglement \cite{Collins:2022akx, Carrozza:2026qcf, Carrozza:2026cci}. In mathematics, tensor models have also recently inspired the birth of free probability for random tensors \cite{bonnin2024freenesstensors, collins2025freecumulantsfreenessunitarily, nechita2025tensorfreeprobabilitytheory}.

A large fraction of the tensor model literature has focused on the large-$N$ limit. In this regime, the perturbative expansion becomes organized by the Gurau degree, melonic diagrams dominate, and many observables satisfy universal factorization properties. This feature was first observed in colored tensor models \cite{Gurau:2009tw,Gurau:2011aq,Bonzom:2011zz,Gurau:2011xq,Gurau:2011xp}, then generalized to other classes of tensor models \cite{Bonzom:2012hw,Gurau:2011kk}. A remarkable consequence is the phenomenon of Gaussian universality, under which the large-$N$ limit of a broad class of tensor models becomes effectively Gaussian: invariant correlation functions are entirely determined by the exact two-point function via Wick's theorem \cite{Gurau:2011xp,Gurau:2011kk}. These developments have led to a detailed understanding of tensor models at large $N$.

By contrast, much less is known about tensor models at finite $N$. Away from the large-$N$ limit, subleading contributions are no longer suppressed, and the simplifying melonic description breaks down. The use of numerics may be necessary to probe this limit, especially in the strong coupling regime\footnote{ By strong coupling, we mean the limit where the coupling related to non-Gaussian corrections of a Gaussian tensor model is taken to be large.}. This motivates the search for methods capable of constraining tensor observables directly, and non-perturbatively, at finite $N$.

In recent years, in margin of the study of tensor models, novel tools have been developed to study matrix models. These tools were first developed for matrix integrals \cite{Lin:2020mme,Kazakov:2021lel}, then matrix quantum mechanics \cite{Han:2020bkb}. Since then, the program has led to new numerical studies of the BFSS matrix model \cite{Banks:1996vh,Lin:2023owt,Lin:2024vvg,Lin:2025srf} (see \cite{Lin:2025iir} for a review), the IKKT matrix model \cite{Ishibashi:1996xs,Li:2025tub}, thermodynamics in the context of matrix quantum mechanics \cite{Cho:2024kxn,Adams:2025nww}, and SUSY matrix quantum mechanics \cite{Laliberte:2025xvk}. For matrix integrals, specifically, the matrix bootstrap has proven useful to constrain the critical behavior of matrix models \cite{Khalkhali:2020jzr}, eigenvalue distributions \cite{Kovacik:2025qgj,Maeta:2026oku}, and unitary matrix integrals \cite{Berenstein:2026wky}. Numerous other directions have also been explored in the literature \cite{Cho:2025vws,Cho:2025nlv,Kazakov:2024ool,Maeta:2026miu}.

As an extension of the matrix bootstrap program, it was recently shown that matrix bootstrap techniques can be used to constrain tensor models that can be mapped onto a rectangular matrix model with skewed scalings \cite{Laliberte:2026qce}. In this work, preliminary bounds were obtained at finite $N$. Following this progress, new positivity statements have been developed to constrain arbitrary unitary tensor integrals \cite{Pagliaroli:2026hxv}. 

The purpose of the present work is to further develop the tensor bootstrap program, and to explore applications in the finite $N$ limit. We introduce two new classes of positivity constraints. The first is constructed from "open bubbles", namely tensor-like objects obtained by removing a tensor from a bubble invariant. The second is constructed from "color matrices" obtained by removing a color contraction from a bubble invariant. These constructions generate additional positive semi-definite conditions that can be combined with Schwinger-Dyson equations to constrain tensor observables. We show that these constraints lead to sharp bounds on tensor integrals at finite $N$ and provide a quantitative probe of deviations from Gaussian universality.

The remainder of this paper is organized as follows. In Sec. \ref{sec:review} we review the basic notions of tensor models relevant for our analysis. In Sec. \ref{sec:positivity} we introduce positivity constraints derived from open bubbles and color matrices. In Sec. \ref{sec:examples} we apply these constraints to finite-$N$ tensor models and study the resulting bounds on bubble observables. We conclude in Sec. \ref{sec:conclusion} with a discussion of future directions for the tensor bootstrap program.

\section{Brief review of tensor models}
\label{sec:review}

Random tensor models are statistical ensembles of tensors whose probability distribution is specified by an action invariant under a prescribed symmetry group. The quantities averaged over in these models are encoded in a real or complex tensor $T_{a_1 a_2 ... a_D}$ with indices $a_i$ ranging from 1 to $N$. The number of indices $D$ is often called the "order" or "rank" of the tensor. As a short-hand notation, we will often write tensors using the vector notation $T_{a_1 a_2 ... a_D} \equiv T_{\vec{a}}$, where $\vec{a} = (a_1 , a_2 , ... , a_D)$ is a vector of indices.

When studying random tensor models, one is usually interested in the properties of observables invariant under this symmetry group, namely bubble invariants, averaged over a predefined ensemble. These invariants are monomials of tensors, where indices are fully summed over, that are invariant under predefined symmetry transformations. 

To illustrate what a bubble invariant is, take for example the quantity
\be
\mathcal{B} = \sum_{a_1 a_2 a_3} \sum_{b_1 b_2 b_3} T_{a_1 a_2 b_3} \bar{T}_{a_1 a_2 a_3} T_{b_1 b_2 a_3} \bar{T}_{b_1 b_2 b_3} \, .
\ee 
This scalar is invariant under the $U(N)^D$ unitary transformations
\begin{align}
	T'_{a_1 \dotsb a_D} = \sum_{b_1 \dotsc b_D} U^{(1)}_{a_1 b_1} \dotsm U^{(D)}_{a_D b_D}\ T_{b_1 \dotsb b_D} \quad , \\
	\bar{T}'_{a_1 \dotsb a_D} = \sum_{b_1 \dotsc b_D} \bar{U}^{(1)}_{a_1 b_1} \dotsm \bar{U}^{(D)}_{a_D b_D}\ \bar{T}_{b_1 \dotsb b_D} \quad ,
\end{align}
and therefore is a bubble invariant for this symmetry group. 

To construct other invariants for the $U(N)^D$ group, one can take any other monomial with an equal number of $T$'s and $\bar{T}$'s, and contract the indices $a_i$ of the $T_{\vec{a}}$'s to the indices $b_j$ of the $\bar{T}_{\vec{b}}$'s for $i=j$. The obtained quantity will be  invariant under the $U(N)^D$ unitary transformations, and therefore a bubble invariant for this group.

Various other groups, such as the orthogonal group $O(N)^D$, have been considered in the study of tensor models. Models that transform under the unitary group $U(N)^D$, defined by the transformations above, will be the focus of our analysis. 

Our goal will be to constrain the expectation value 
\be
\left\langle \mathcal{B}(T,\bar{T}) \right\rangle \equiv \frac1{Z} \int [dT\,d\bar{T}]\ \mathcal{B}(T,\bar{T})\ e^{-N^{D-1}\,S(T,\bar{T})} \,
\ee
of bubbles transforming under the unitary group for a tensor model with the potential $S(T , \bar{T})$. We will be interested in models with a partition function $Z$ and tensor potential $S(T , \bar{T})$ of the following form
\begin{align}
	Z = \int [dT\,d\bar{T}]\ e^{-N^{D-1}\,S(T,\bar{T})} \label{eq:Z} \quad , \\
	S(T,\bar{T}) = T\cdot \bar{T} + \sum_{i\in I} t_i\,\mathcal{B}_i(T,\bar{T}) \quad .
\end{align}
Here, $T \cdot \bar{T} = \sum_{\vec{a}} T_{\vec{a}} \bar{T}_{\vec{a}}$ is a Gaussian term and the $\mathcal{B}_i(T,\bar{T})$'s are bubble invariants with an associated coupling constant $t_i$.  Below, we review some notions of unitary tensor models that will be helpful for our analysis.

\noindent
\textbf{Schwinger-Dyson Equations:} The Schwinger-Dyson equations are consistency conditions obtained by imposing that tensor integrals are invariant under a change of variables. Consider, for example, the partition function $Z$. Imposing invariance under the infinitesimal transformations $T_{\vec{a}} \rightarrow T_{\vec{a}} + \epsilon \, \mathcal{O}_{\vec{a}}$ yields the condition
\be
\frac{1}{Z} \sum_{\vec{a}} \int dT d\bar{T} \frac{\partial}{\partial T_{\vec{a}}} \left( \mathcal{O}_{\vec{a}} \, e^{-N^{D-1} S(T , \bar{T})}\right) = 0 \, .\footnote{See \cite{DiFrancesco:1993cyw} for an example of this derivation in the matrix case.}
\label{eq:sde1}
\ee
Here, we will assume that $\mathcal{O}_{\vec{a}}$ is a quantity that transforms like $T_{\vec{a}}$ under unitary transformations. Similarly, one could consider imposing that $Z$ is invariant under the transformation $\bar{T}_{\vec{a}} \rightarrow \bar{T}_{\vec{a}} + \epsilon \, \bar{\mathcal{O}}_{\vec{a}}$. Here, we assume that $ \bar{\mathcal{O}}_{\vec{a}}$ is a quantity that transforms like $\bar{T}$. In this case, we obtain
\be
\frac{1}{Z} \sum_{\vec{a}} \int dT d\bar{T} \frac{\partial}{\partial \bar{T}_{\vec{a}}} \left( \bar{\mathcal{O}}_{\vec{a}} \, e^{-N^{D-1} S(T , \bar{T})}\right) = 0 \, .
\label{eq:sde2}
\ee
Another way to motivate the identities (\ref{eq:sde1}) and (\ref{eq:sde2}) is to notice that they describe the integral of a total derivative. Therefore, they should trivially be zero. Evaluating both expressions, one obtains the simple identities
\begin{align}
\sum_{\vec{a}}  \left\langle \frac{\partial \mathcal{O}_{\vec{a}} }{\partial T_{\vec{a}}} \right\rangle =  \sum_{\vec{a}}  \left\langle N^{D-1} \, \mathcal{O}_{\vec{a}} \frac{\partial S(T , \bar{T})}{\partial T_{\vec{a}}} \right\rangle \label{eq:sde_white}
\quad , \\
\sum_{\vec{a}}  \left\langle \frac{\partial \bar{\mathcal{O}}_{\vec{a}} }{\partial \bar{T}_{\vec{a}}} \right\rangle =  \sum_{\vec{a}}  \left\langle N^{D-1} \, \bar{\mathcal{O}}_{\vec{a}} \frac{\partial S(T , \bar{T})}{\partial \bar{T}_{\vec{a}}} \right\rangle \quad . \label{eq:sde_black}
\end{align}
Making use of the identities above, one can relate the expectation value $\langle \mathcal{B} \rangle$ of the bubbles of a tensor model using a linear set of equations (see \cite{Gurau:2012ix,Bonzom:2012hw,Bonzom:2012cu} for relevant studies). These equations will be used as part of our analysis to constrain the possible values of $\langle \mathcal{B} \rangle$.

\noindent
\textbf{Tensors as graphs:} In the study of tensor models, one usually encounters long monomials of tensors with complicated ways of contracting indices. To simplify notation and analysis, it is common in the literature to circumvent this problem by describing tensors as graphs. 

For complex tensors transforming under the unitary group, tensors $T_{\vec{a}}$ are often represented by white vertices with open edges describing indices, while tensors $\bar{T}_{\vec{a}}$ are often represented by black vertices with open edges describing indices (see an example for rank 3 tensors below).
\begin{equation*}
	T_{a_1 a_2 a_3} = \graphimage{1.25cm}{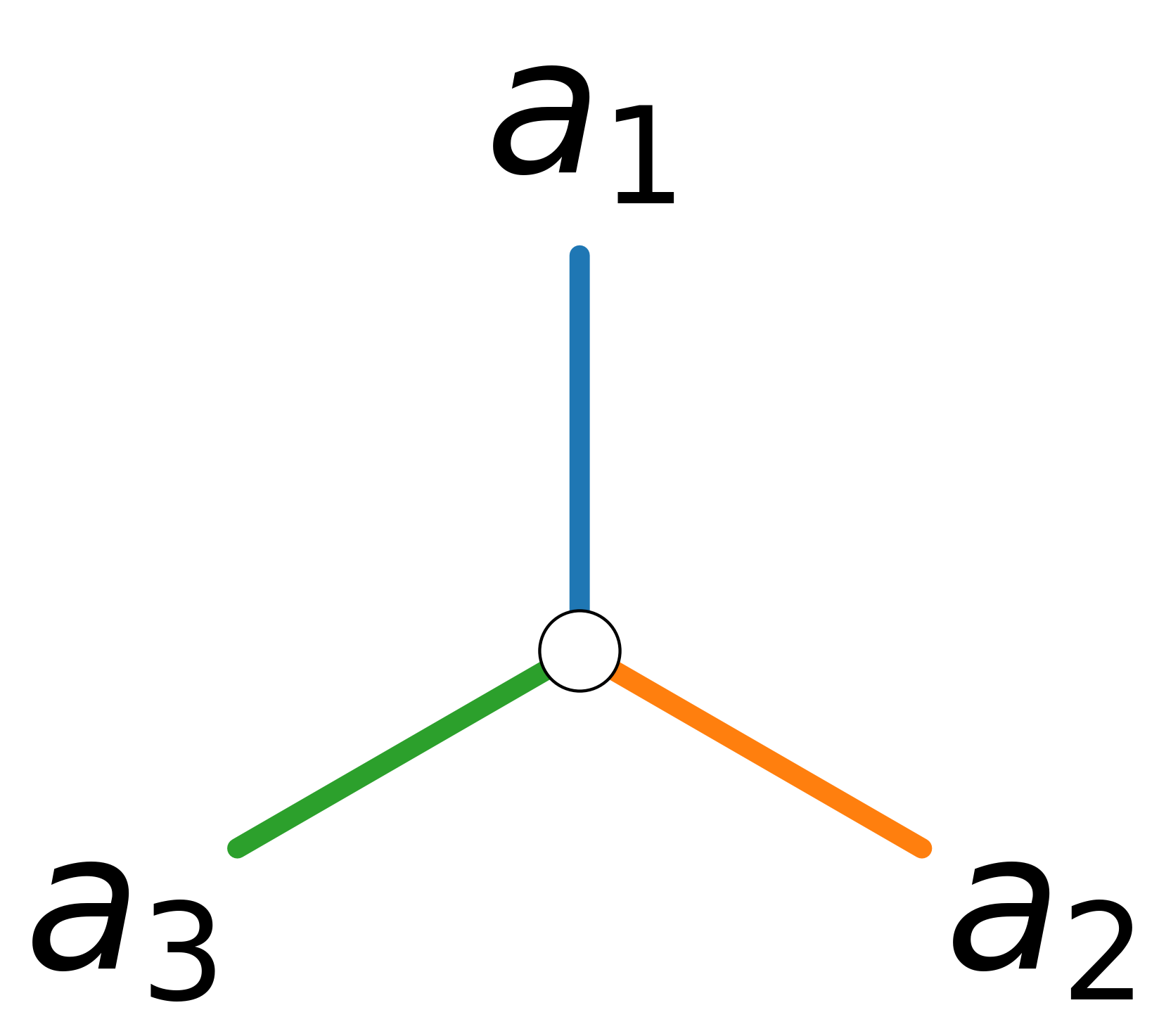}
	\quad , \quad
	\bar{T}_{a_1 a_2 a_3} = \graphimage{1.25cm}{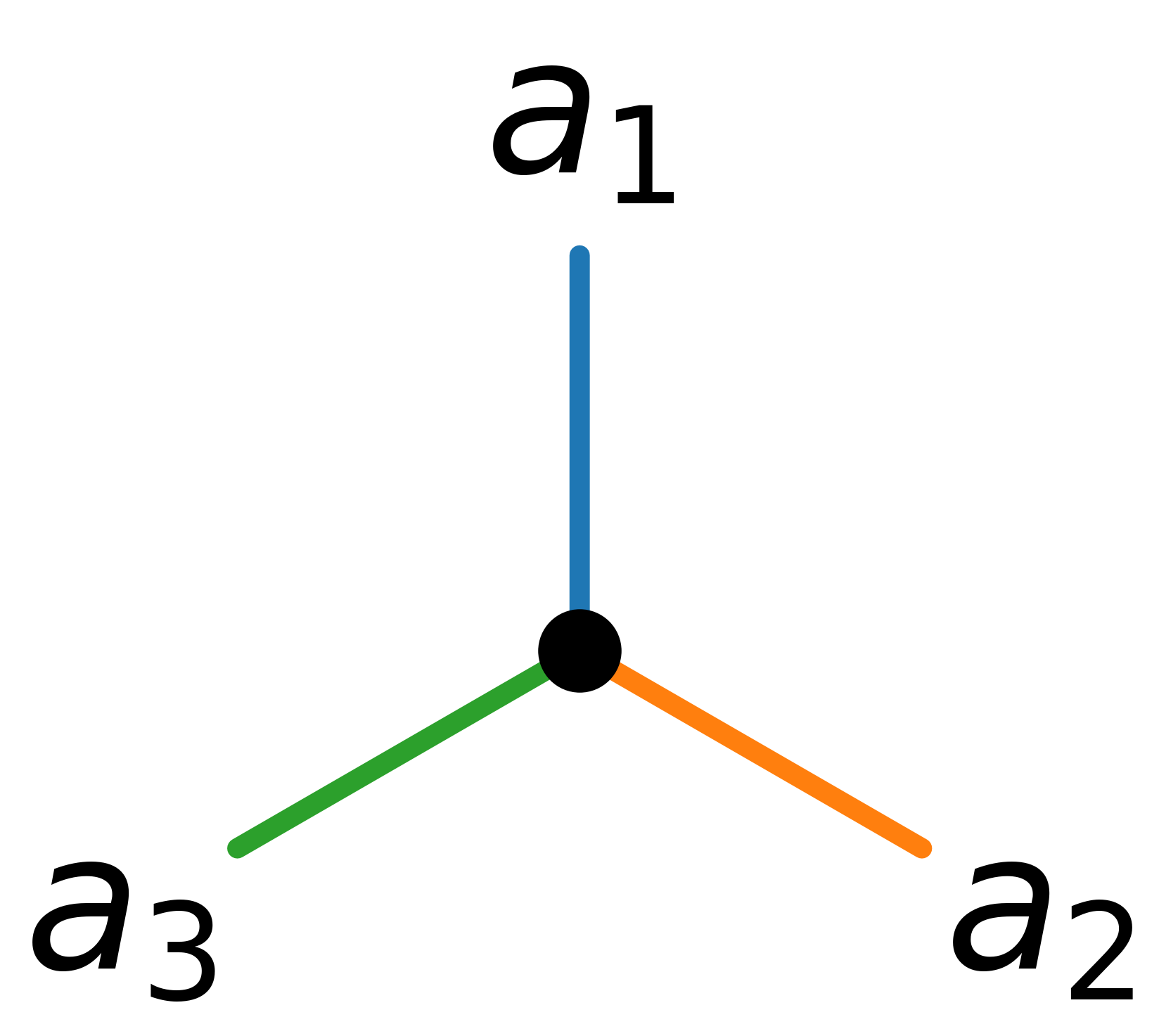} \quad .
\end{equation*}
This is known as the bipartite graph notation. One can then create bubble invariants by fully connecting edges of white vertices to edges of black vertices with matching color (see, again, examples for a rank 3 tensor below).
\begin{align*}
\sum_{a_1 a_2 a_3} T_{a_1 a_2 a_3} \bar{T}_{a_1 a_2 a_3} & = \graphimage{1.25cm}{bubbles/bubble_rank3_n1_001.png} \quad , \\ 
\sum_{a_1 a_2 a_3} \sum_{b_1 b_2 b_3} T_{a_1 a_2 b_3} \bar{T}_{a_1 a_2 a_3} T_{b_1 b_2 a_3} \bar{T}_{b_1 b_2 b_3} & = \graphimage{1.25cm}{bubbles/bubble_rank3_n2_001.png} \quad , \\ 
... 
\end{align*}
In the present case, white vertices must be connected to black vertices to preserve unitary invariance.

\noindent
\textbf{Melonic graphs and Gaussian universality:} For Gaussian tensor models with unitary invariance, there is a special set of graphs named melonic graphs. These graphs are generated recursively from the elementary melon $T \cdot \bar{T}$ by repeated melonic insertions.

A melonic insertion of color $i$ consists of replacing a color-$i$ contraction between a tensor $T$ and a conjugate tensor $\bar{T}$ by the following melonic piece
\begin{equation}
M^{(i)}_{a_i b_i}
=
\sum_{\substack{a_1 \ldots a_{i-1}\\a_{i+1} \ldots a_D}}
T_{a_1 \cdots a_{i-1} a_i a_{i+1} \cdots a_D}
\bar{T}_{a_1 \cdots a_{i-1} b_i a_{i+1} \cdots a_D}.
\end{equation}

Graphically, one removes a color-$i$ edge and replaces it by two new vertices connected by $D-1$ colored edges, leaving two external half-edges of color $i$. Repeating this operation on any color-$i$ edge of an existing melonic graph generates the full family of melonic observables.

To illustrate this process, let us consider the elementary melon $\sum_{a_1 a_2 a_3} T_{a_1 a_2 a_3} \bar{T}_{a_1 a_2 a_3}$ for a rank 3 tensor. In this case, there are three possible melonic pieces. These take the form
\begin{equation}
\begin{gathered}
M^{(1)}_{a_1 b_1} = \colormatrices{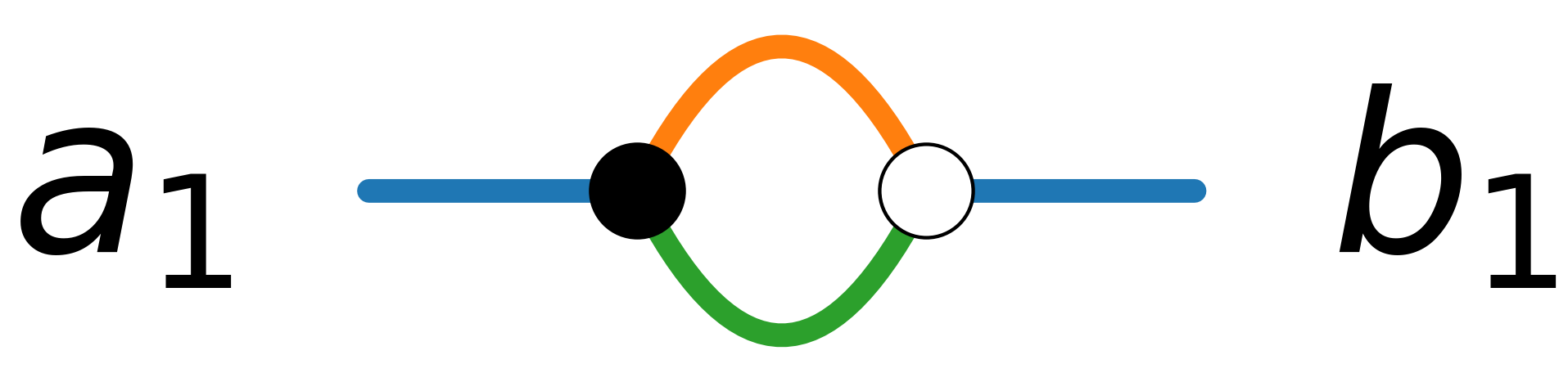}
\quad , \quad
M^{(2)}_{a_2 b_2} = \colormatrices{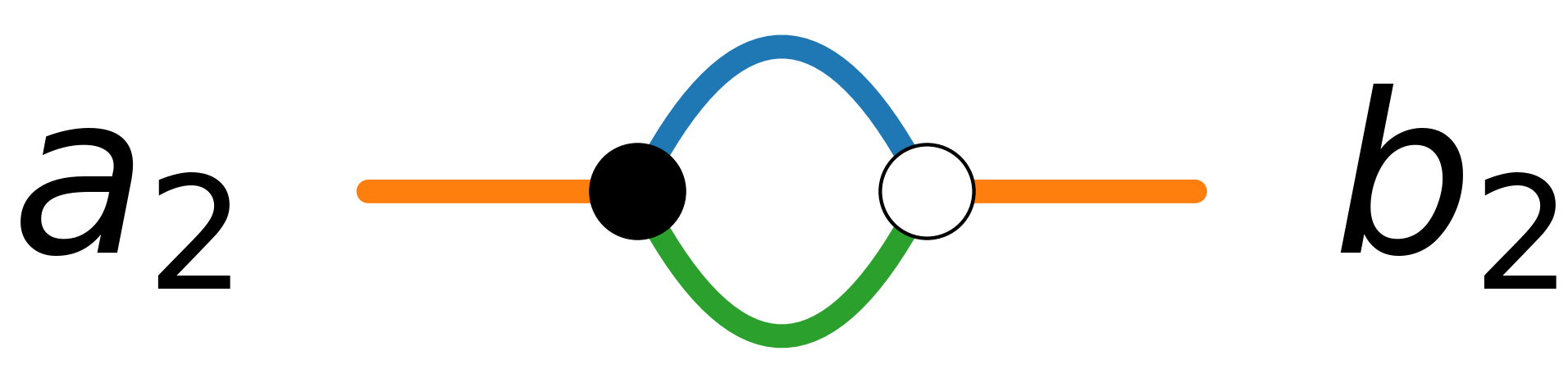}
\quad , \quad \\
M^{(3)}_{a_3 b_3} = \colormatrices{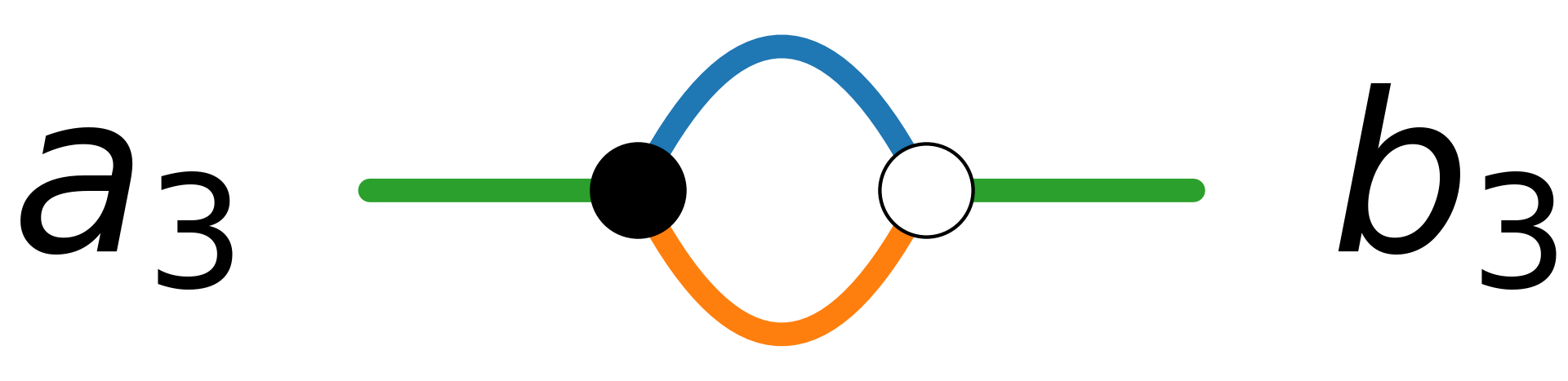}
\quad .
\end{gathered}
\end{equation}
As an example, let us consider a melonic insertion on the blue edge of the elementary melon for rank 3 tensors. In this case, we obtain
\begin{equation*}
\raisebox{-0.5\height}{\includegraphics[width=1.25cm]{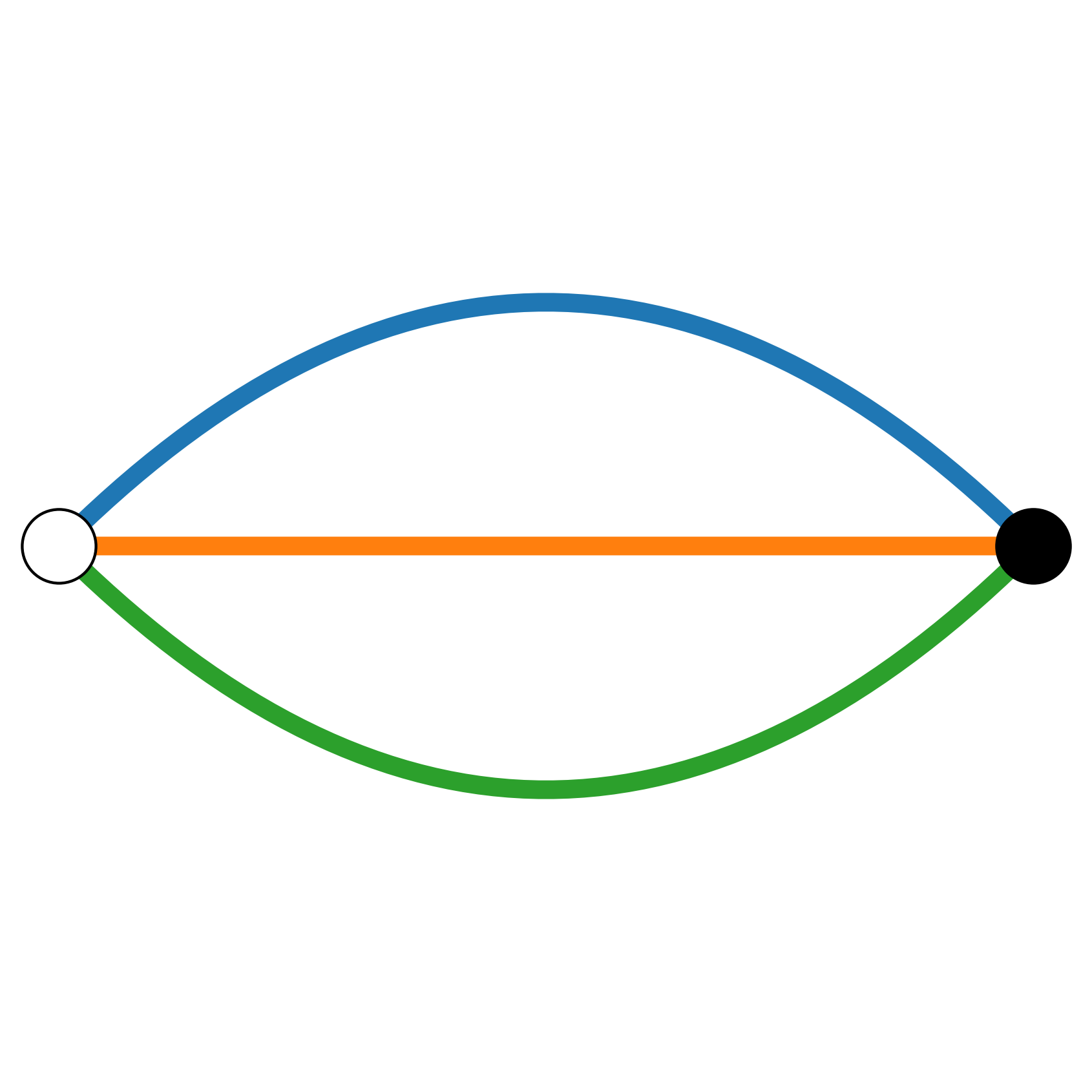}} 
\quad \rightarrow \quad 
\raisebox{-0.5\height}{\includegraphics[width=1.25cm]{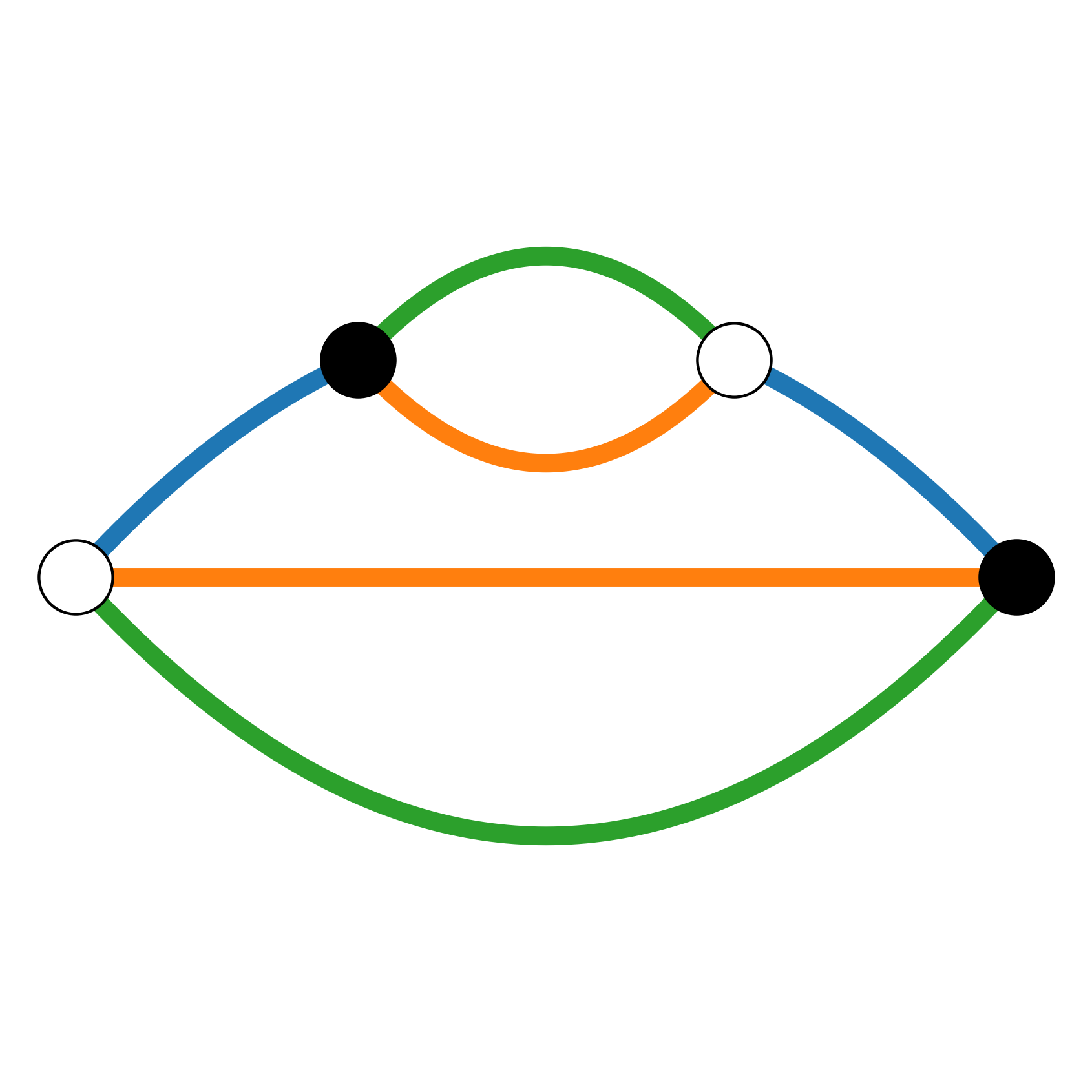}} \quad .
\end{equation*}
From repeated melonic insertion on any other edge of the elementary melon, one can obtain all the other melonic graphs for the present system. 

Melonic graphs have the special property that they dominate the free energy $F = -\log N^{D-1} Z$ of tensor models in the large $N$ limit \cite{Gurau:2011aq,Bonzom:2011zz,Gurau:2011xq,Bonzom:2012hw}. In particular, in this limit, they obey the well-known equation
\be
\left\langle \frac{1}{N} \mathcal{B}_M \right\rangle = \left\langle \frac{1}{N} \melon \right\rangle^{v/2} \, .
\label{eq:gauss_univ}
\ee
Here, $v$ is the number of vertices in the melonic bubble $\mathcal{B}_M$. This property of melonic graphs in the large $N$ limit is called Gaussian universality \cite{Bonzom:2012hw,Gurau:2011kk}.

Gaussian universality can make tensor models remarkably easy to solve in the large $N$ limit. To illustrate this fact, let us consider a tensor model with the potential
\be
S = \melon + g \left( \, \pillowone + \, \pillowtwo \right) \, .
\label{eq:two_pillow_int}
\ee
To solve for the elementary melon at large $N$, one can consider the following Schwinger-Dyson equation
\be
\frac{1}{Z} \sum_{\vec{a}} \int dT d\bar{T} \frac{\partial}{\partial T_{\vec{a}}} \left( T_{\vec{a}} \, e^{-N^{D-1} S(T , \bar{T})}\right) = 0 .
\ee
In this case, taking the derivative and dividing by a factor of $N^D$ lets us recover the following expression
\be
1 = \left\langle \frac{1}{N} \melon \right\rangle + 2 g \left\langle \frac{1}{N} \pillowone \right\rangle + 2 g \left\langle \frac{1}{N} \pillowtwo \right\rangle \, .
\label{eq:sde_two_pillow_system}
\ee
In the large $N$ limit, one expects the expectation values of the pillows to be related to the expectation values of the elementary melon via the expression
\be
\left\langle \frac{1}{N} \pillowone \right\rangle = \left\langle \frac{1}{N} \pillowtwo \right\rangle = \left\langle \frac{1}{N} \melon \right\rangle^2 \, . 
\ee
This follows directly from Gaussian universality \eqref{eq:gauss_univ}. Making use of this identity, \eqref{eq:sde_two_pillow_system} becomes a polynomial and can be solved exactly. We obtain
\be
\left\langle \frac{1}{N} \pillowone \right\rangle = \frac{- 1 + \sqrt{1+16 g}}{8 g} \, .
\ee
for the expectation value of the elementary melon at large $N$. Note that the equation above generalizes nicely to a Gaussian model that has a potential with an arbitrary number $n$ of quartic pillows. In this case, the large $N$ result becomes
\be
\left\langle \frac{1}{N} \pillowone \right\rangle = \frac{- 1 + \sqrt{ 1 + 8 g n}}{4 g n} \, .
\label{eq:melon_largeNexact}
\ee

At finite $N$, the Schwinger-Dyson equations of a given system can become much more complicated to solve. In this regime, observables do not factorize and one must consider a large number of linear equations. Generally, the exact solution of these equations is not known. This motivates us to define additional consistency conditions, such as positivity, to constrain the set of potential solutions at finite $N$.

\section{Positivity constraints}
\label{sec:positivity}

To constrain tensor integrals, we will make use of positivity constraints constructed from tensor-like and matrix-like quantities. To derive these constraints, we will start from the observation that the statement
\be
\sum_{\vec{a}} \langle \Phi_{\vec{a}} \, \bar{\Phi}_{\vec{a}} \rangle \geq 0 \, ,
\ee
must be true for any tensor $\Phi_{\vec{a}}$ under the assumption of a positive probability measure. Using the statement above, one can then imagine constructing $\Phi_{\vec{a}}$ using a sum of tensors $\mathcal{O}_{\vec{a}}$ such that
\be
\Phi_{\vec{a}} = \sum_i c^i \mathcal{O}^i_{\vec{a}} \, .
\ee
Here, the $c^i$'s are arbitrary complex coefficients. In this case, the inequality above reads
\be
\sum_{\vec{a}} \langle \Phi_{\vec{a}} \bar{\Phi}_{\vec{a}} \rangle = \sum_{\vec{a}} \sum_{ij} \bar{c}^i \langle \mathcal{O}^i_{\vec{a}} \bar{\mathcal{O}}^j_{\vec{a}} \rangle c^j  \geq 0 \, .
\label{eq:positivity_constraint}
\ee
We obtain that the Gram matrix
\be
\mathcal{M}_{ij} = \sum_{\vec{a}}  \langle \mathcal{O}^i_{\vec{a}} \,  \bar {\mathcal{O}}^j_{\vec{a}} \rangle \,
\label{eq:gram_mat}
\ee 
must be positive semi-definite $(\mathcal{M} \succeq 0)$. 

It is easy to see that the above quantity is a generalization of the positivity statement used in the matrix. Let us consider the case where $D = 2$. In this case, the Gram matrix above reads
\be
\mathcal{M}_{ij} = \sum_{a b}  \langle \mathcal{O}^i_{a b} \,  \bar{\mathcal{O}}^j_{a b} \rangle = \langle \Tr \big(\mathcal{O}^i (\mathcal{O}^j)^\dagger\big) \rangle \, .
\ee
This is the statement widely used to constrain matrix models using positivity.

In contrast with the matrix case, there are many classes of tensor and matrix-like structures that one can use in a Gram matrix of the form (\ref{eq:gram_mat}). This, in some sense, makes the study of the tensor bootstrap potentially much richer than the study of the matrix bootstrap. In the present analysis, we will restrict ourselves to Gram matrices made of two types of tensor and matrix-like observables: "open bubbles" and "color matrices". There are, of course, many other structures one could consider. However, the ones mentioned here will prove sufficient to obtain strong bounds for the classes of models that we will study. We go over the construction of these Gram matrices below.

\begin{figure}[h]
	\centering
	\includegraphics[width=\linewidth]{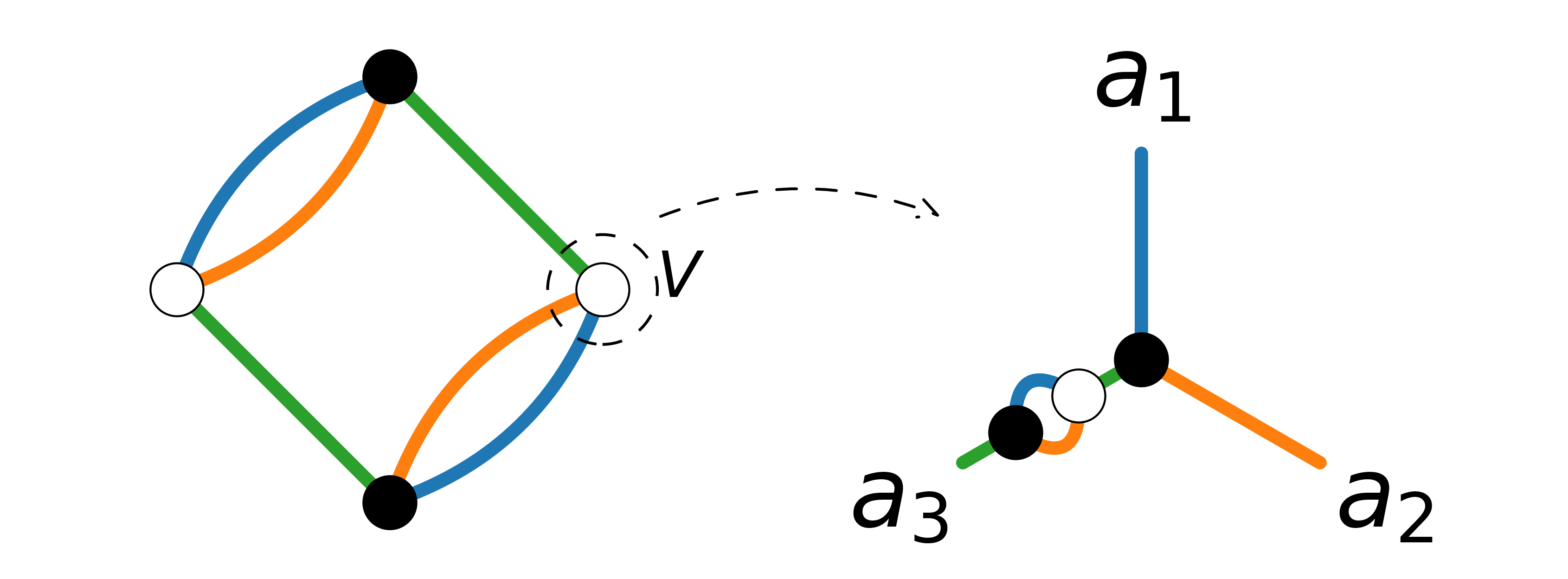}
	\caption{Schematic representation of an open bubble. When the vertex $V$ is cut and removed from the bubble $\mathcal{B}$ on the left, one obtains a tensorial object $(\mathcal{B} \, \backslash \, V)_{a_1 a_2 a_3}$, which transforms like a tensor $T_{a_1 a_2 a_3}$ under unitary transformations.}
	\label{fig:open_bubble}
\end{figure}

\noindent
\textbf{Gram matrix for open bubbles:} The first Gram matrix we will use is one made of open bubbles. Given a bubble invariant $\mathcal{B}$, one can define an open bubble $\mathcal{B} \, \backslash \, V$ as the tensor-like quantity obtained from removing a white vertex V from a bubble invariant (see Figure \ref{fig:open_bubble} for an example). The remaining quantity, which has tensor elements $(\mathcal{B} \, \backslash \, V)_{\vec{a}}$, then transforms as a $\bar{T}$. One could also choose to remove a black vertex $\bar{V}$ instead. In this case, the open bubble $\mathcal{B} \, \backslash \, \bar{V}$ transforms as a $T$.

To construct the matrix, we will consider a vector of $\mathcal{O}^i_{a_1  a_2  a_3}$ of open bubbles found from opening all bubble invariants up to $2n$ vertices, where $n$ is the number of black/white vertex pairs found in the invariant. For a rank 3 tensor, opening all bubbles with 2 vertices, 4 vertices, and so forth gives a vector of the form below
\be
\mathcal{O}^i_{a_1  a_2  a_3} = \left( \eqimage{open_bubbles/open_bubble_removed_black_n1_001.png} \, , \, \eqimage{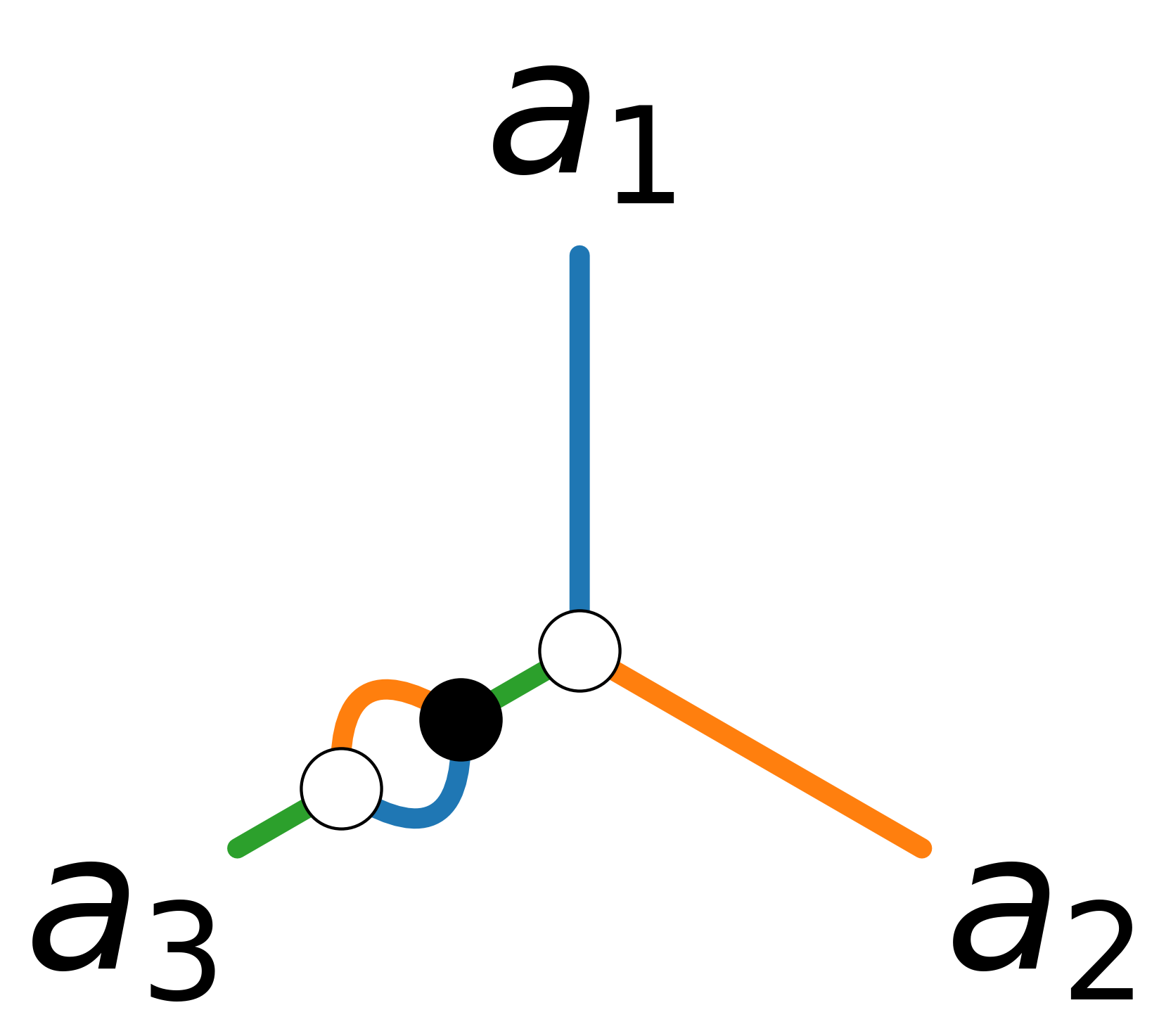} \, , \, \eqimage{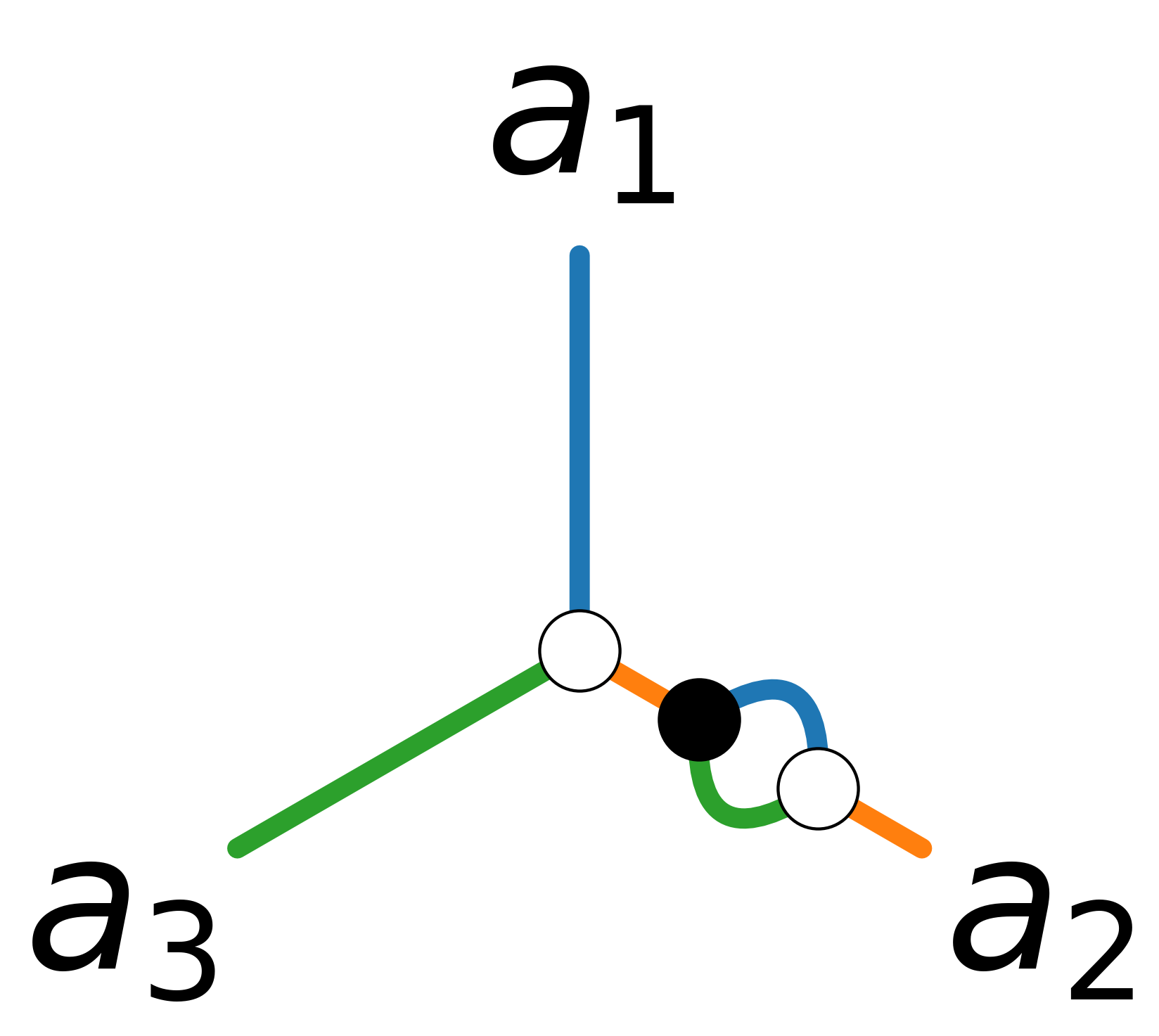} \, , \, \eqimage{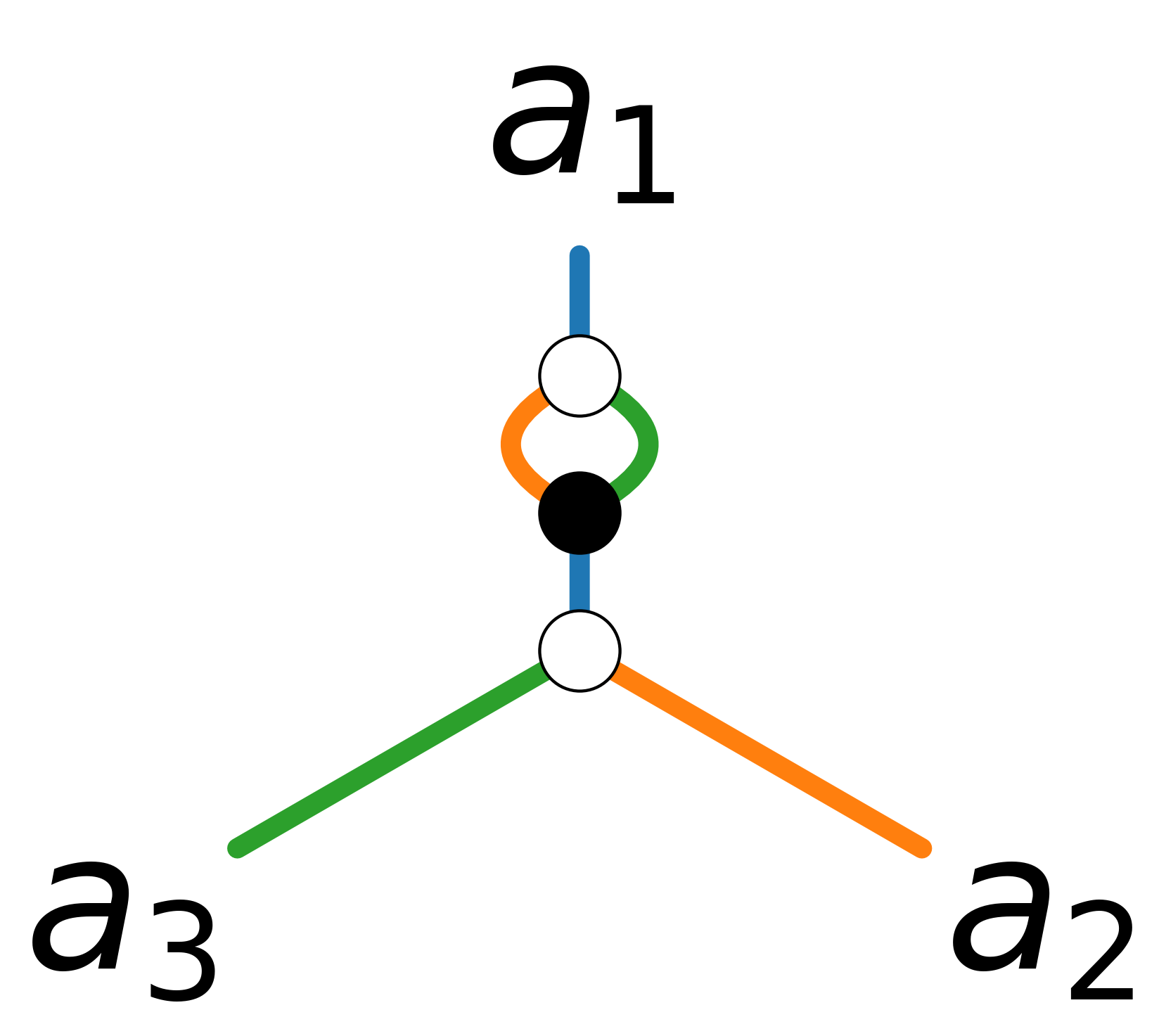} \, ,  ... \right) \, .
\ee
Contracting two such vectors, one then obtains a Gram matrix of the form below
\be
\mathcal{N}_{ij} = \left(\raisebox{-0.5\height}{\includegraphics[width=5cm]{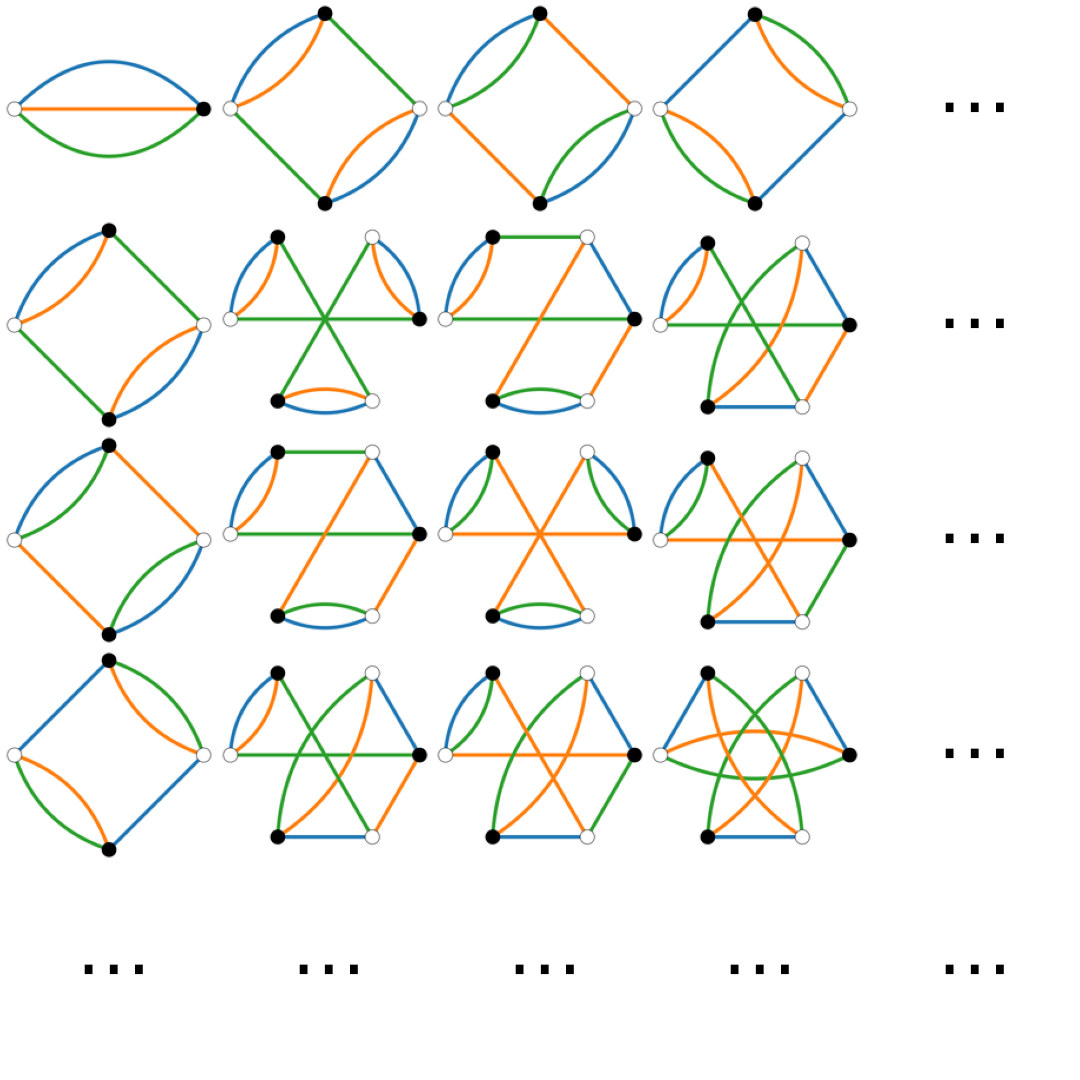}}\right) \, .
\ee
In the above, we adopted the convention
\be
\mathcal{M}_{ij} =  \left\langle \frac{1}{N} \mathcal{N}_{ij} \right\rangle \, \quad , \quad \mathcal{N}_{ij} = \sum_{\vec{a}} \bar{\mathcal{O}}^i_{\vec{a}} \,  \mathcal{O}^j_{\vec{a}}
\ee 
for our Gram matrices. This is done so that expectation values of bubble invariants in the Gram matrices are weighted by a factor $1/N$. In the tensor model literature, this $1/N$ scaling is often introduced so that expectation values of melonic bubbles return $\mathcal{O}(1)$ quantities. We will adopt this convention for the expectation value of all bubbles in our analysis.

\begin{figure}[h]
	\centering
	\includegraphics[width=\linewidth]{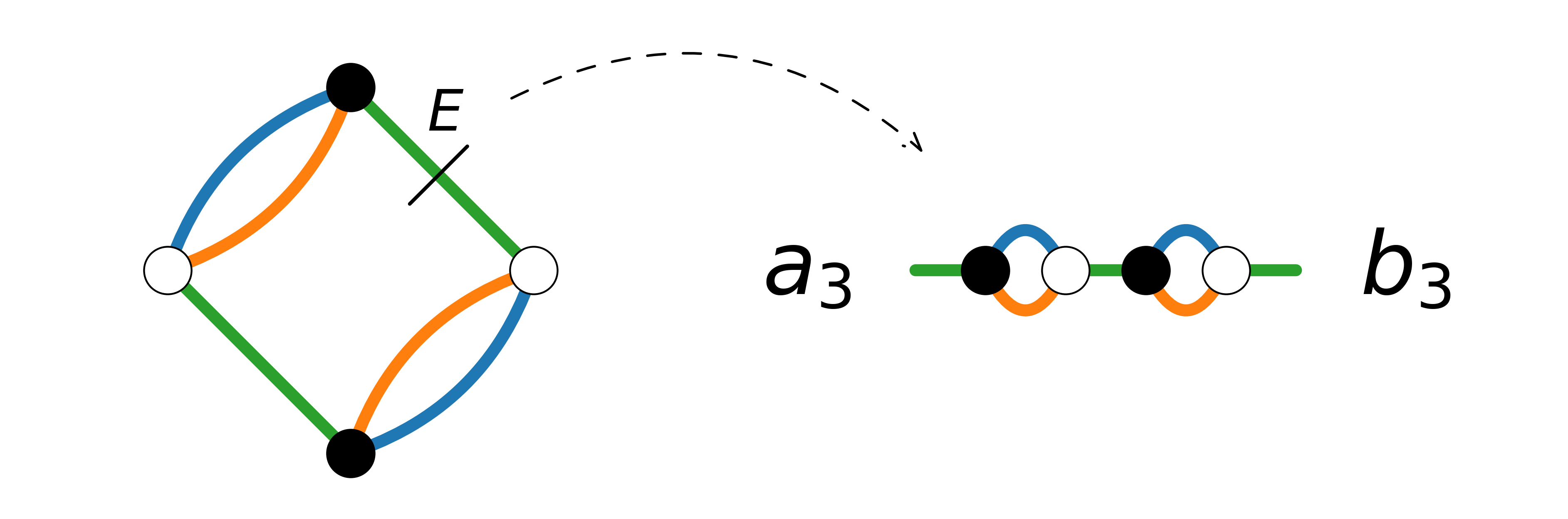}
	\caption{Schematic representation of a color matrix. When an edge $E$ is cut open from a bubble invariant, one obtains a matrix with two open color indices $a_i$ and $b_i$. In the image above, a green edge is cut from a bubble invariant $\mathcal{B}$, leading to the color matrix $(\mathcal{B} \, \backslash \, E)_{a_3 b_3}$.}
	\label{fig:color_matrix}
\end{figure}

\noindent
\textbf{Gram matrix for color matrices:} The second type of Gram matrix we will use is one made of color matrices. Given a bubble invariant $\mathcal{B}$, one can define a color matrix $\mathcal{B} \, \backslash \, E$ as the matrix-like quantity obtained from cutting an edge $E$ related to a color contraction in a bipartite graph (see Figure \ref{fig:color_matrix} for an example). The remaining quantity, which has matrix elements $(\mathcal{B} \, \backslash \, E)_{a_i b_i}$, transforms as a $U(N)$ matrix for the color $i$ related to the cut edge.

To construct the Gram matrix, we will again consider a vector of open graphs. However, this time, the vector $\mathcal{O}^i_{a_i  b_i}$ will be made of color matrices found from cutting open an edge from each bubble invariant containing up to $2n$ vertices, where $n$ is the number of black/white vertex pairs in the invariant. For a rank 3 tensor, cutting up all bubbles with 2 vertices, 4 vertices, and so forth for the green color gives a vector of the form below
\begin{align}
\mathcal{O}^i_{a_3  b_3} = \left( \colormatrices{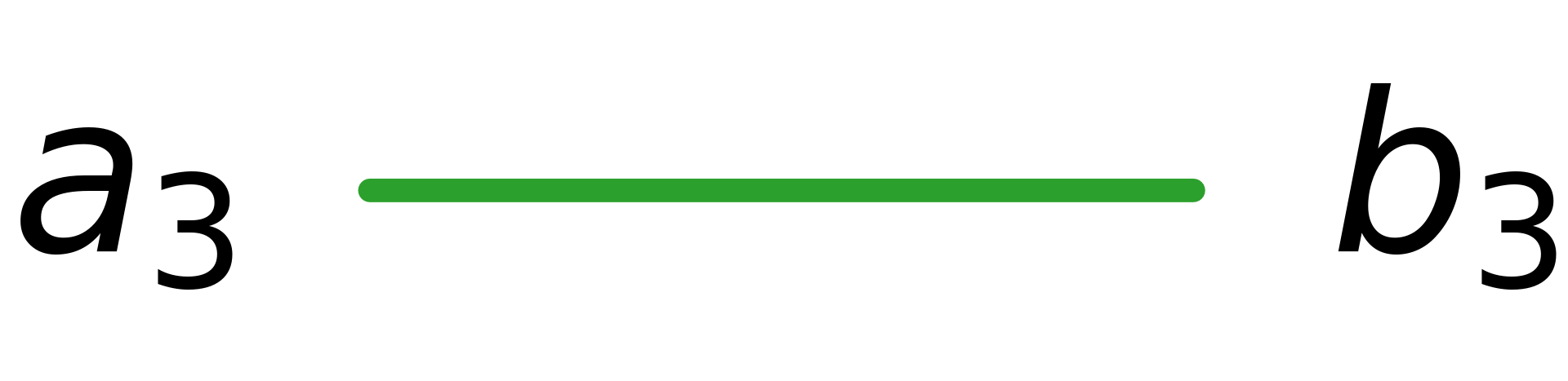} \, , \, \colormatrices{open_graph_n1_color2_001.png} \, , \, \colormatrices{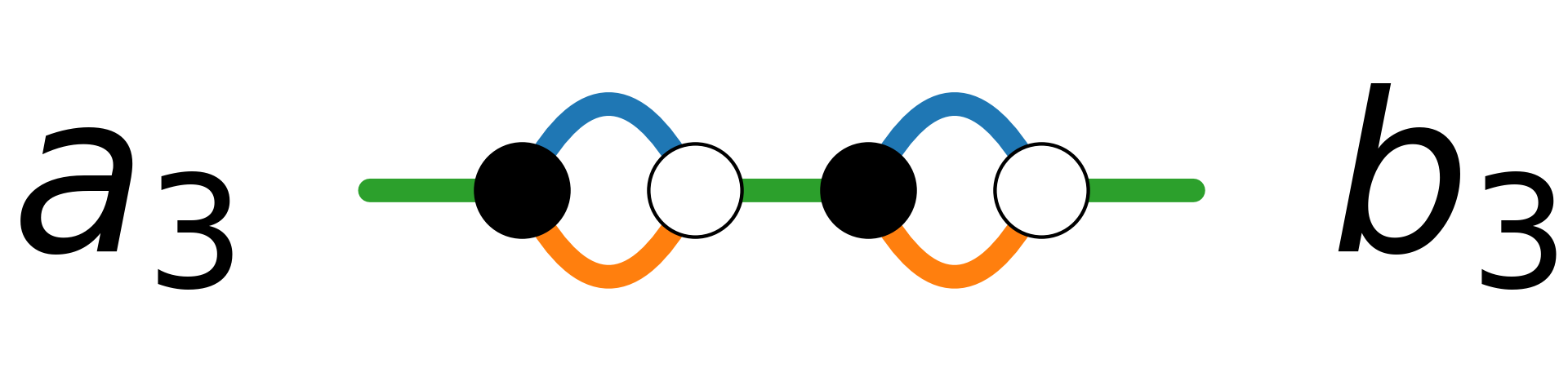} \, , \, \right. \nonumber \\ \left. \colormatrices{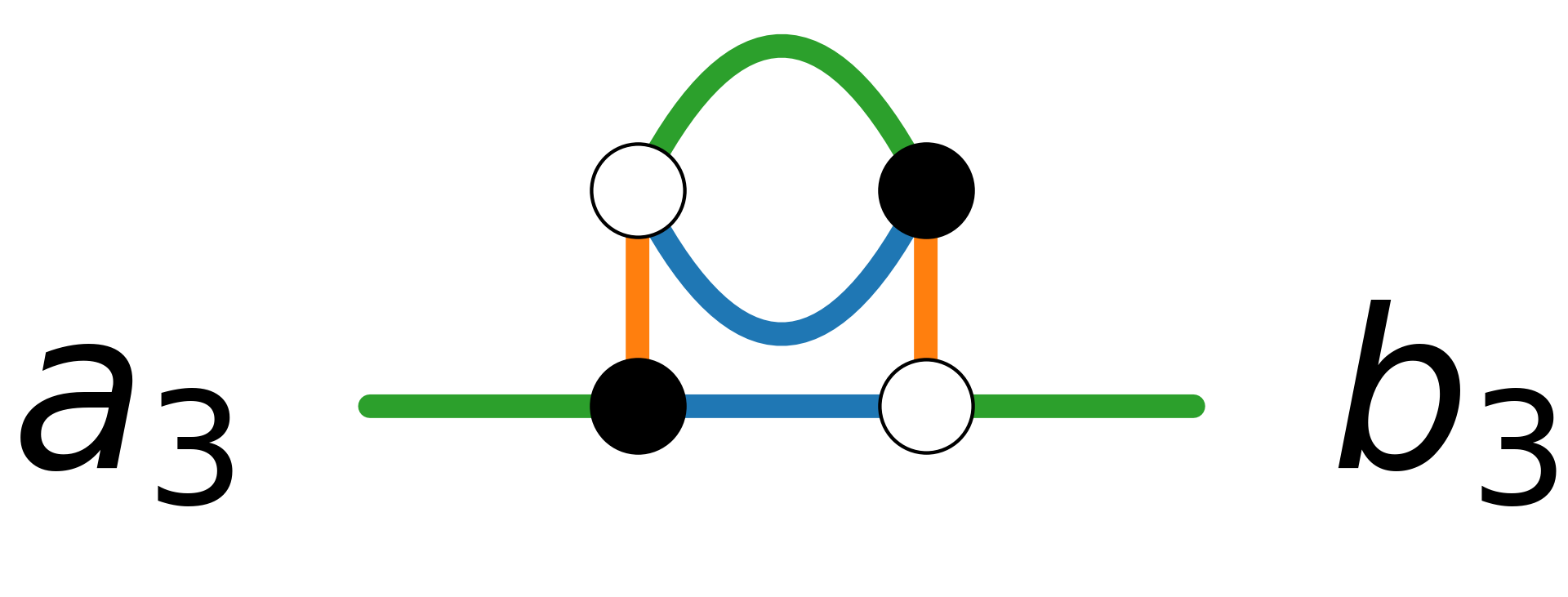} \, , \, \colormatrices{open_graph_n2_color2_002.png} \, , \,
... \right) 
\label{eq:color_vector}
\end{align}
For convenience, we have also added the Kronecker delta function $\delta_{a_3 b_3} = \colormatrices{open_graph_n0_color2_001.png}$ as first element of this vector. For any color matrix  $(\mathcal{B} \, \backslash \, E)_{a_i b_i}$, contraction with the Kronecker delta lets us recover the original invariant $\mathcal{B} = \sum_{a_i b_i} (\mathcal{B} \, \backslash \, E)_{a_i b_i} \delta_{a_i b_i}$. Moreover, the Kronecker delta transforms unitarily under the $U(N)$ group. Therefore, it is a perfectly valid entry. 

Contracting two color matrices of the form (\ref{eq:color_vector}), one obtains a Gram matrix of the form below
\be
\mathcal{N}_{ij} = \left(\raisebox{-0.5\height}{\includegraphics[width=5cm]{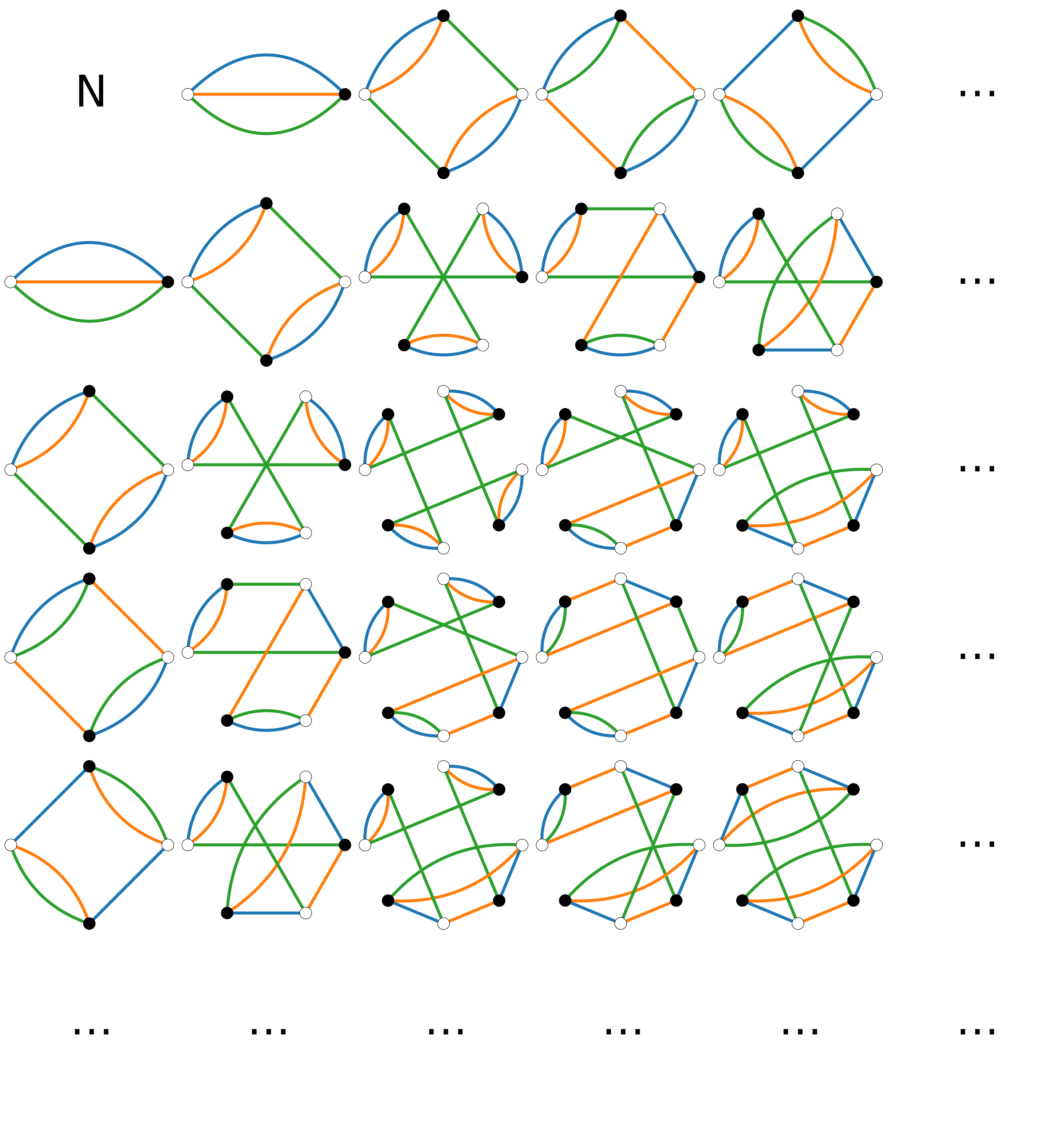}}\right)
\ee
Given (\ref{eq:positivity_constraint}), the open bubble Gram matrix and the color Gram matrices must be positive semi-definite as long as the model has a positive measure. This allows for bounds on the bubble invariants of any system with such a measure.

\section{Examples}
\label{sec:examples}

Let us see some examples of how positivity can be used to obtain sharp bounds on tensor integrals at finite $N$, and study deviations from Gaussian universality. For this purpose, we will use positivity of open bubbles, color matrices, and the Schwinger-Dyson equations of a given system. Our approach is as follows. 

\textbf{Step 1:} We generate the Gram matrices found from removing a vertex (open bubble case) or an edge (color matrix) from bubble invariants up to $2n$ vertices, where $n$ is the number of black/white vertex pairs in the invariant.

\textbf{Step 2:} We generate all Schwinger-Dyson equations associated with open bubbles whose resulting equations involve only bubbles with no more vertices than those appearing in the Gram matrices\footnote{Note that we only used open bubbles to generate the Schwinger-Dyson for our analysis. Color matrices only have 2 indices and do not transform like a tensor $T_{\vec{a}}$. Therefore, they are not valid objects to consider for this step.}. 

\textbf{Step 3:} We partially solve the Schwinger-Dyson equations by relating bubbles with a larger number of vertices to those with a smaller number of vertices, and substituting the expression for bubbles with a larger number of vertices in the Gram matrices.

\textbf{Step 4:} We determine upper and lower bounds on the bubble expectation value $\langle \mathcal{B} \rangle$ by solving optimization problems that maximize and minimize $\langle \mathcal{B} \rangle$ under the requirement that the Gram matrices remain positive semi-definite.

Before diving into a specific example, we will add a few clarifications regarding the implementation of our approach.

For a system described by a rank $D$ tensor, there is one open bubble Gram matrix and $D$ color matrix Gram matrices. Each of these $(D+1)$ Gram matrices is generated.

To generate all Schwinger-Dyson equations, we use the open bubbles found by removing a white vertex and the open bubbles found by removing a black vertex. These equations can be computed explicitly from (\ref{eq:sde_white}) and (\ref{eq:sde_black}).

To proceed with the extremization step, we make use of semi-definite programming, treating each bubble expectation value as a variable to optimize over. Several good algorithms are readily available online for this task. In the present case, we made use of SDPA \cite{Yamashita2012SDPA}, which we found had the best balance between accuracy and compute time.

In the present analysis, the number of vertex pairs $n$ effectively acts as the truncation "level" of the constraints. The larger $n$ is, the stronger the constraints are. However, this comes at the cost of having to generate more bubble invariants and longer compute time.

\noindent
\textbf{Example 1 - Three pillow interaction:} Let us now illustrate our approach using a tensor model with the potential
\be
S = \melon + g \left( \, \pillowone + \, \pillowtwo + \, \pillowthree \right) \, ,
\label{eq:three_pillow_int}
\ee
as an example. The interaction term in this potential is often referred to as a three-pillow interaction.

We first generate the Gram matrices found from opening and cutting open bubble invariants containing up to 6 vertices, or in other words 3 black/white vertex pairs. For the present example, there are 4 such Gram matrices. The first one, made from open bubbles, has $17 \times 17$ matrix elements. The other three, made from color matrices, have $18 \times 18$ matrix elements. The first few entries of the open bubble Gram matrix have been displayed in the previous section, and can be found in Figure \ref{fig:open_bubble}. Similarly, the first few entries of the Gram matrix found by cutting open the green color of bubble invariants can be found in Figure \ref{fig:color_matrix}.

\begin{figure*}[t]
	\centering
	\includegraphics[height=1.0cm]{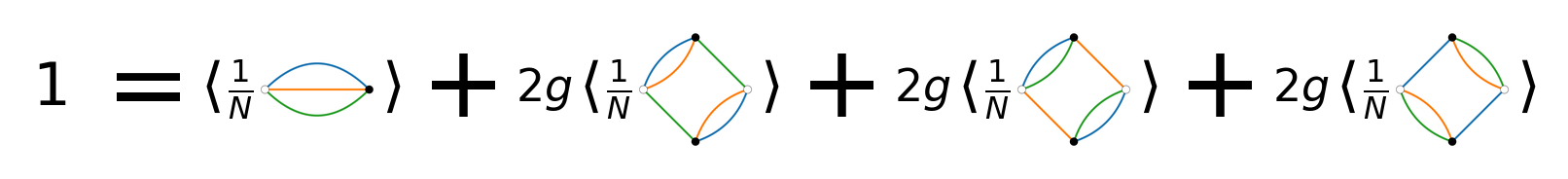}\\
	\includegraphics[height=1.0cm]{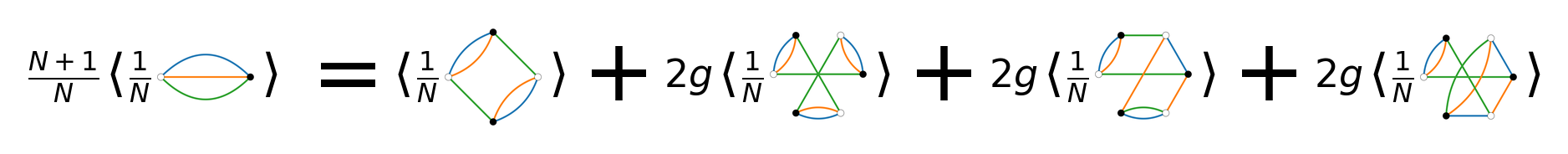}\\
	\includegraphics[height=1.0cm]{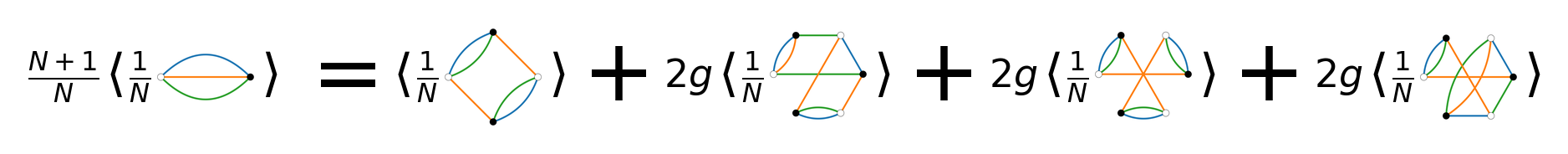}\\
	\includegraphics[height=1.0cm]{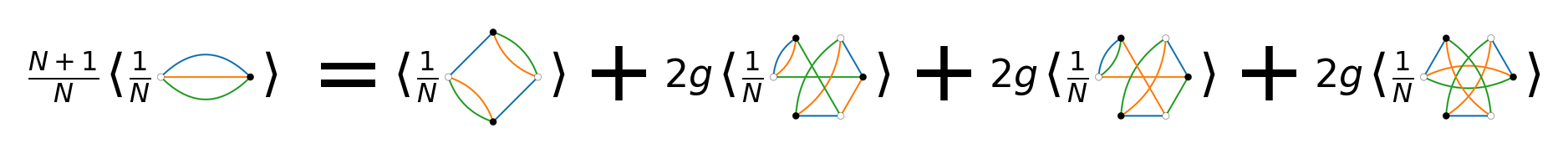}\\
	...
	\caption{Above, we enumerate the first four Schwinger-Dyson equations for the three-bubble system.}
	\label{fig:three_pillows_sdes}
\end{figure*}

We then generate the Schwinger-Dyson equations for the present system. In the Gram matrices for the present system, there are bubble invariants that contain up to 12 vertices, or in other words 6 black/white vertex pairs. This means we must generate all Schwinger-Dyson equations that contain bubbles of at most 12 vertices. For the present system, we found that the set of open bubbles found from opening invariants containing up to 10 vertices, or 5 black/white vertex pairs, generated all relevant equations. Below, we present the first few elements of this set using two different sets describing the values that the operators $\mathcal{O}_{a_1 a_2 a_3}$ and $\bar{\mathcal{O}}_{a_1 a_2 a_3}$ can take in the Schwinger-Dyson equations (\ref{eq:sde_white}) and (\ref{eq:sde_black}).
\begin{align}
\mathcal{O}_{a_1 a_2 a_3}
& \in
\left\{
\eqimage{open_bubbles/open_bubble_removed_black_n1_001.png} \, , \,
\eqimage{open_bubbles/open_bubble_removed_black_n2_001.png} \, , \,
\eqimage{open_bubbles/open_bubble_removed_black_n2_002.png} \, , \,
\eqimage{open_bubbles/open_bubble_removed_black_n2_003.png} \, , \, ...
\right\} \\
\bar{\mathcal{O}}_{a_1 a_2 a_3}
& \in
\left\{
\eqimage{open_bubbles/open_bubble_removed_white_n1_001.png} \, , \,
\eqimage{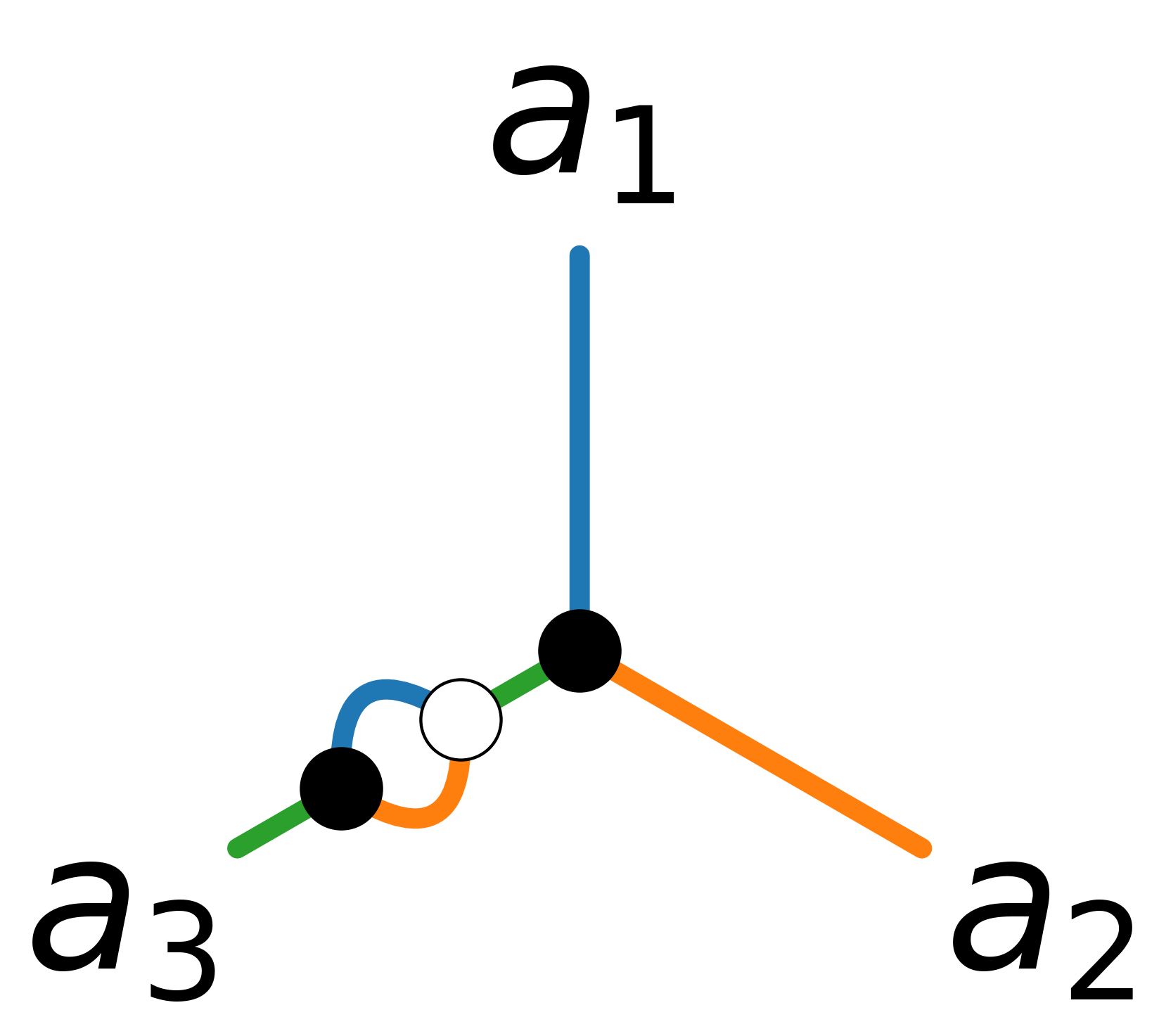} \, , \,
\eqimage{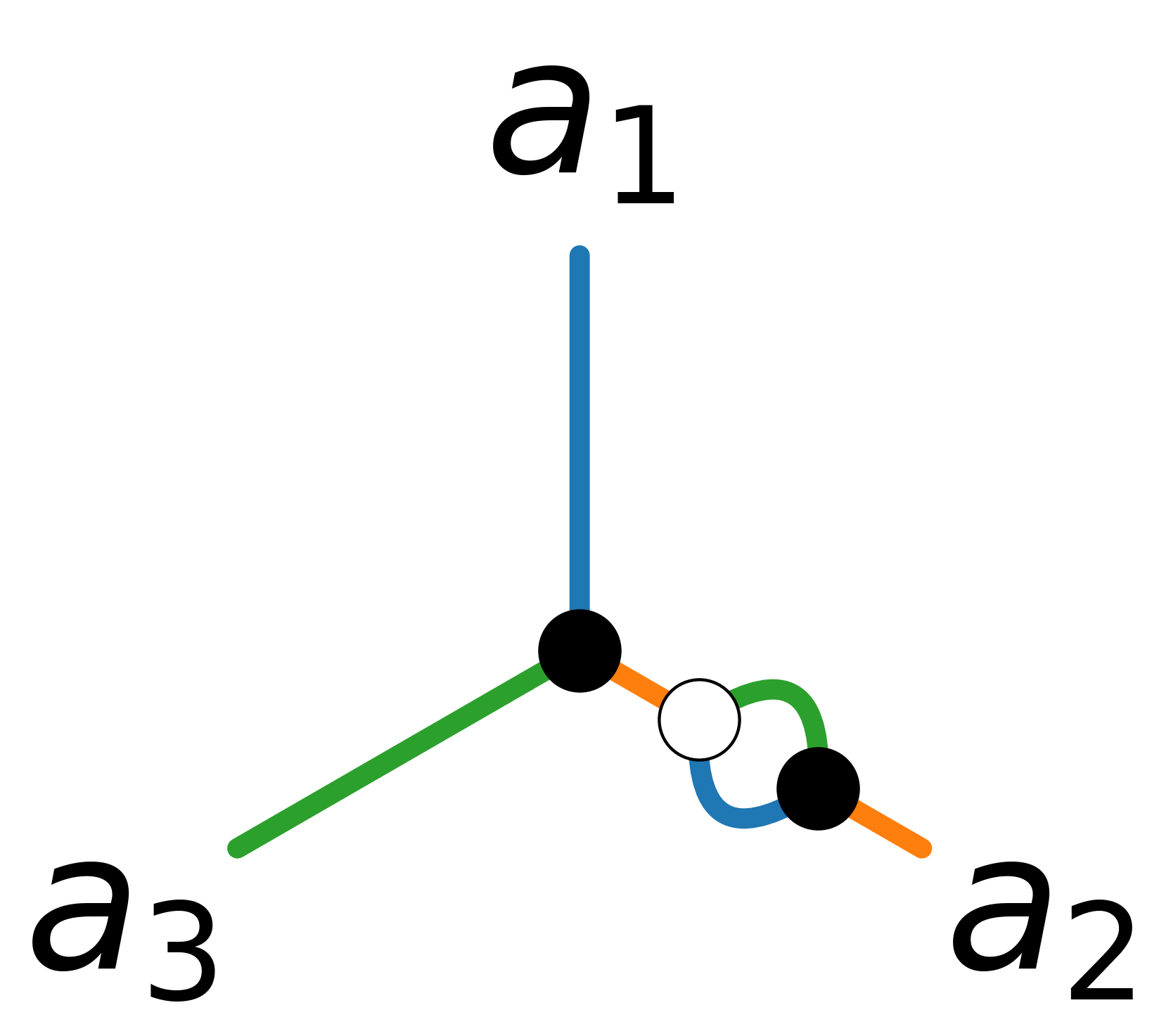} \, , \,
\eqimage{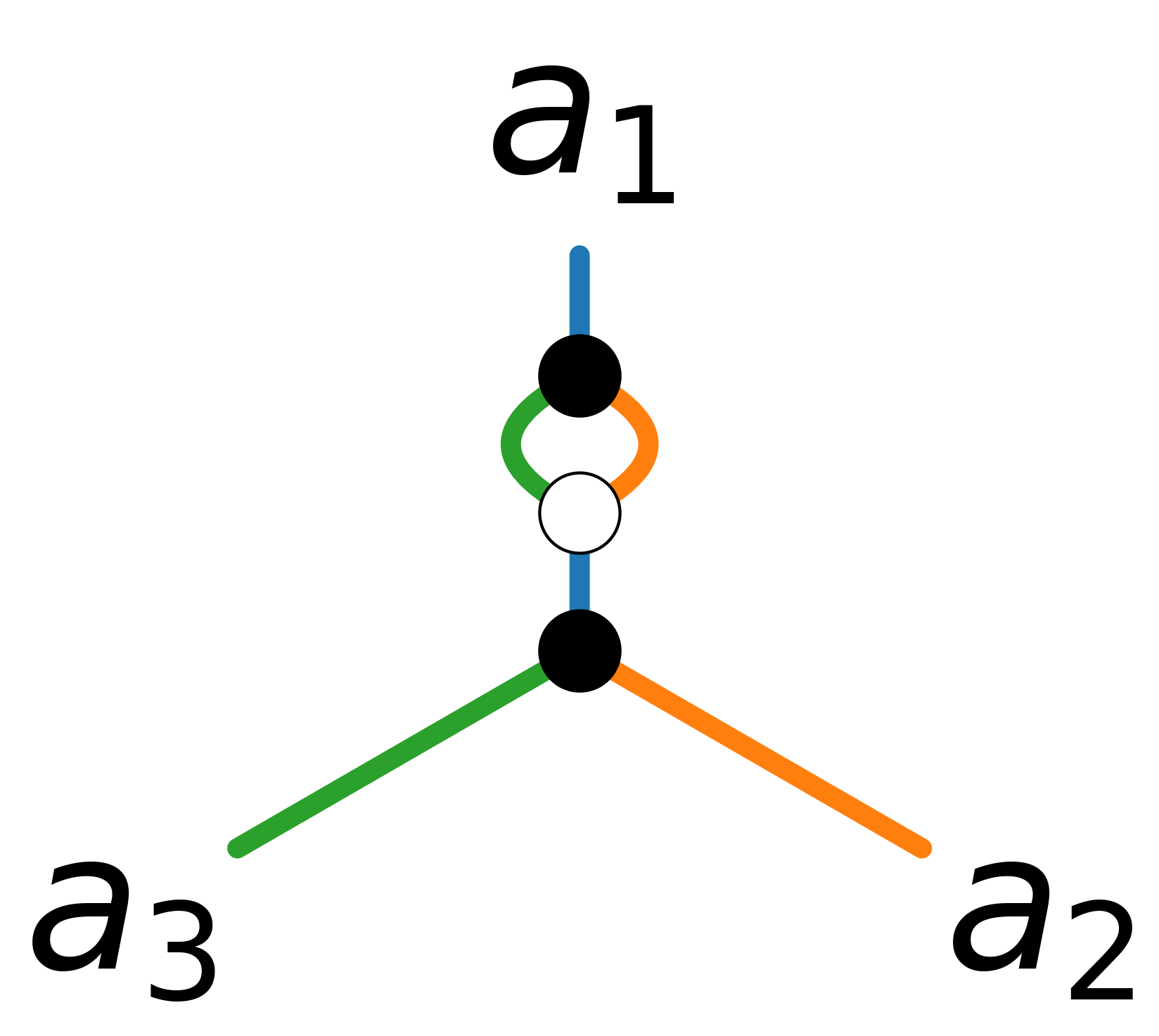} \, , \, ...
\right\}
\end{align}
In the above, $\mathcal{O}_{a_1 a_2 a_3}$ is an open bubble found by removing a black vertex from an invariant, and $\bar{\mathcal{O}}_{a_1 a_2 a_3}$ is an open bubble found by removing a white vertex from an invariant. The first few equations found by inserting the above values of $\mathcal{O}_{a_1 a_2 a_3}$ and $\bar{\mathcal{O}}_{a_1 a_2 a_3}$ in the Schwinger-Dyson equations (\ref{eq:sde_white}) and (\ref{eq:sde_black}) can be found in Figure \ref{fig:three_pillows_sdes}. Some of these equations are, of course, redundant. However, we computed them for the sake of completeness.

We then partially solve the Schwinger-Dyson equations for the present system. This is done by relating bubbles with a larger number of vertices to bubbles with a lower number of vertices. Once the system of equations is partially solved, we substitute the solved variables in the Gram matrices themselves in preparation for the semi-definite programming step.

Finally, we find the maximum and minimum allowed values of an expectation value $\langle \mathcal{B} \rangle$ of a bubble $\mathcal{B}$ for the present system under the constraint that the Gram matrices must be positive semi-definite. This is done by fixing the value of the coupling $g$ in the Gram matrices, and finding the maximum and minimum allowed value of this bubble using semi-definite programming. This process is repeated for various values of $g$ within a predetermined interval to obtain bounds on the expectation value as a function of $g$.

\begin{figure*}[t]
    \centering

    \begin{minipage}{0.45\textwidth}
        \centering
        \includegraphics[width=\linewidth]{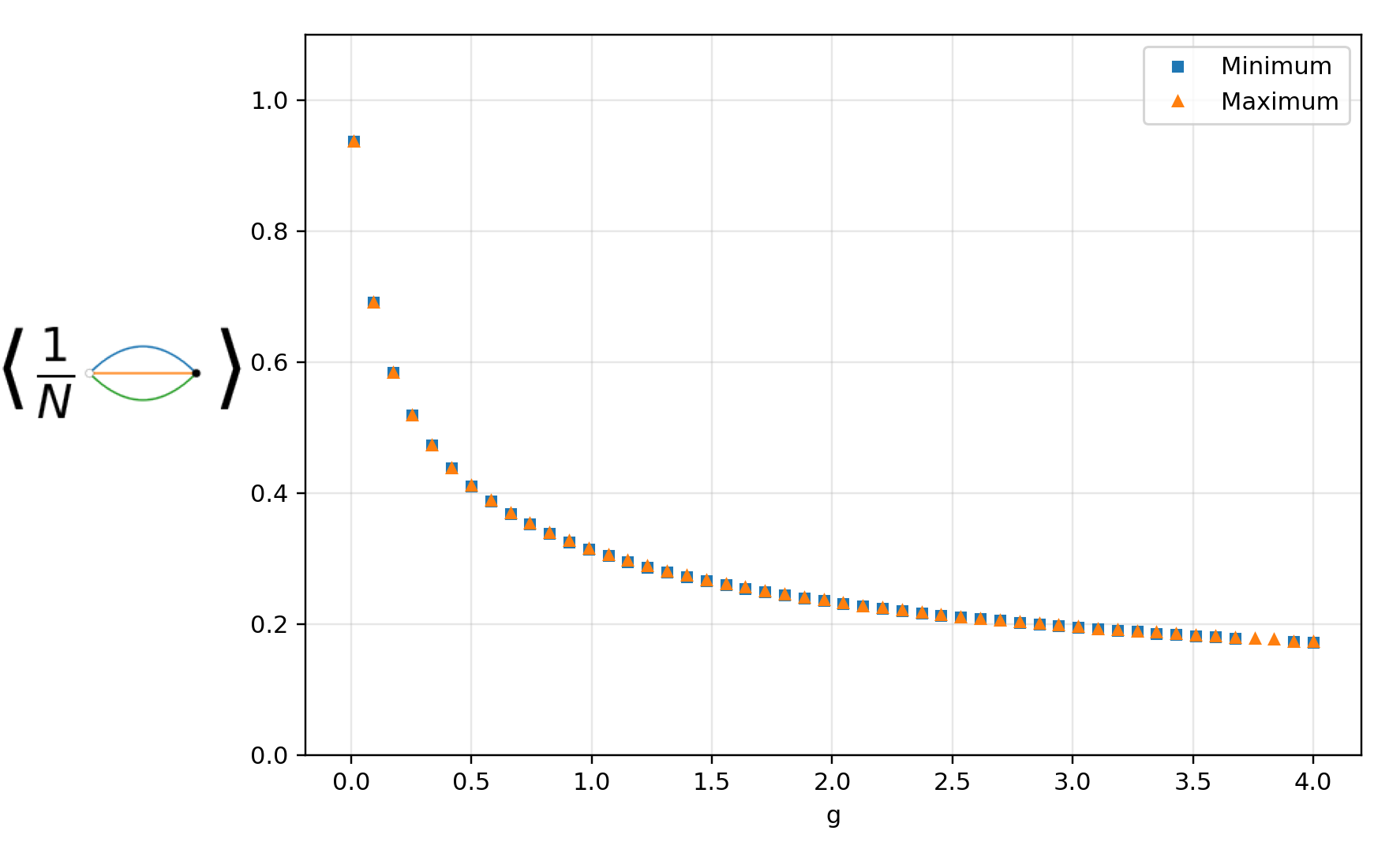}
    \end{minipage}
    \hspace{0.03\textwidth}
    \begin{minipage}{0.45\textwidth}
        \centering
        \includegraphics[width=\linewidth]{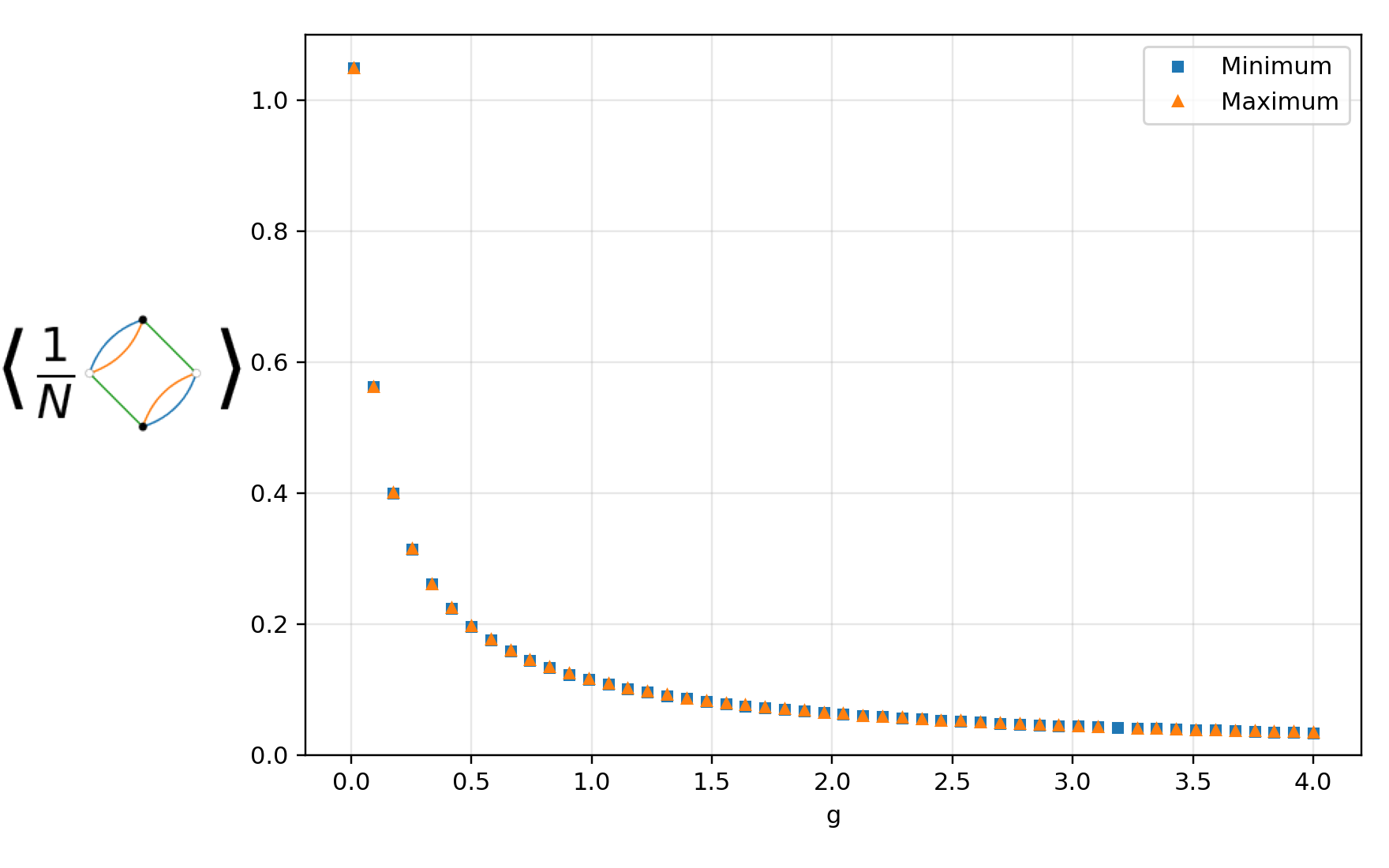}
    \end{minipage}

    \vspace{0.5cm}

    \begin{minipage}{0.45\textwidth}
        \centering
        \includegraphics[width=\linewidth]{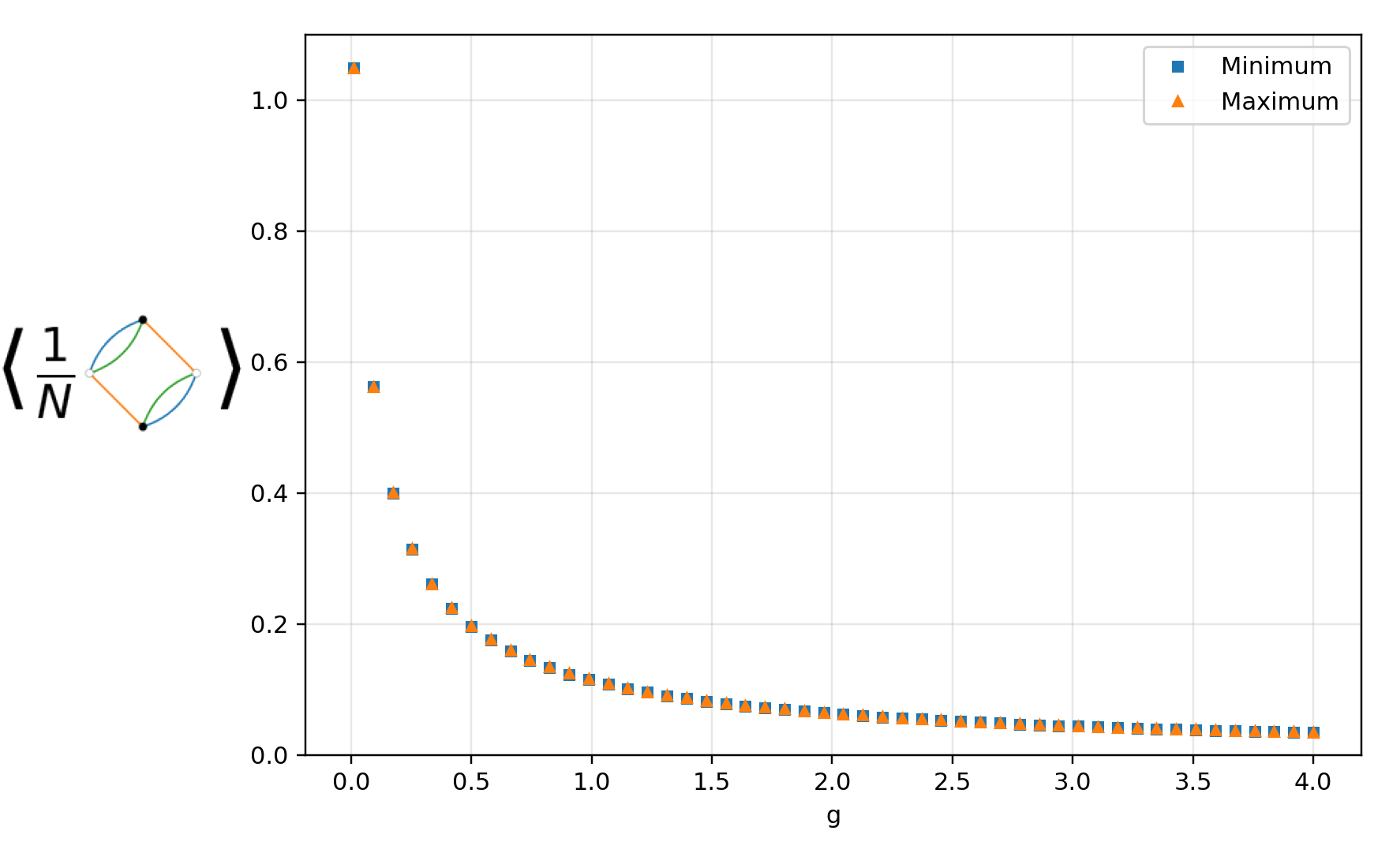}
    \end{minipage}
    \hspace{0.03\textwidth}
    \begin{minipage}{0.45\textwidth}
        \centering
        \includegraphics[width=\linewidth]{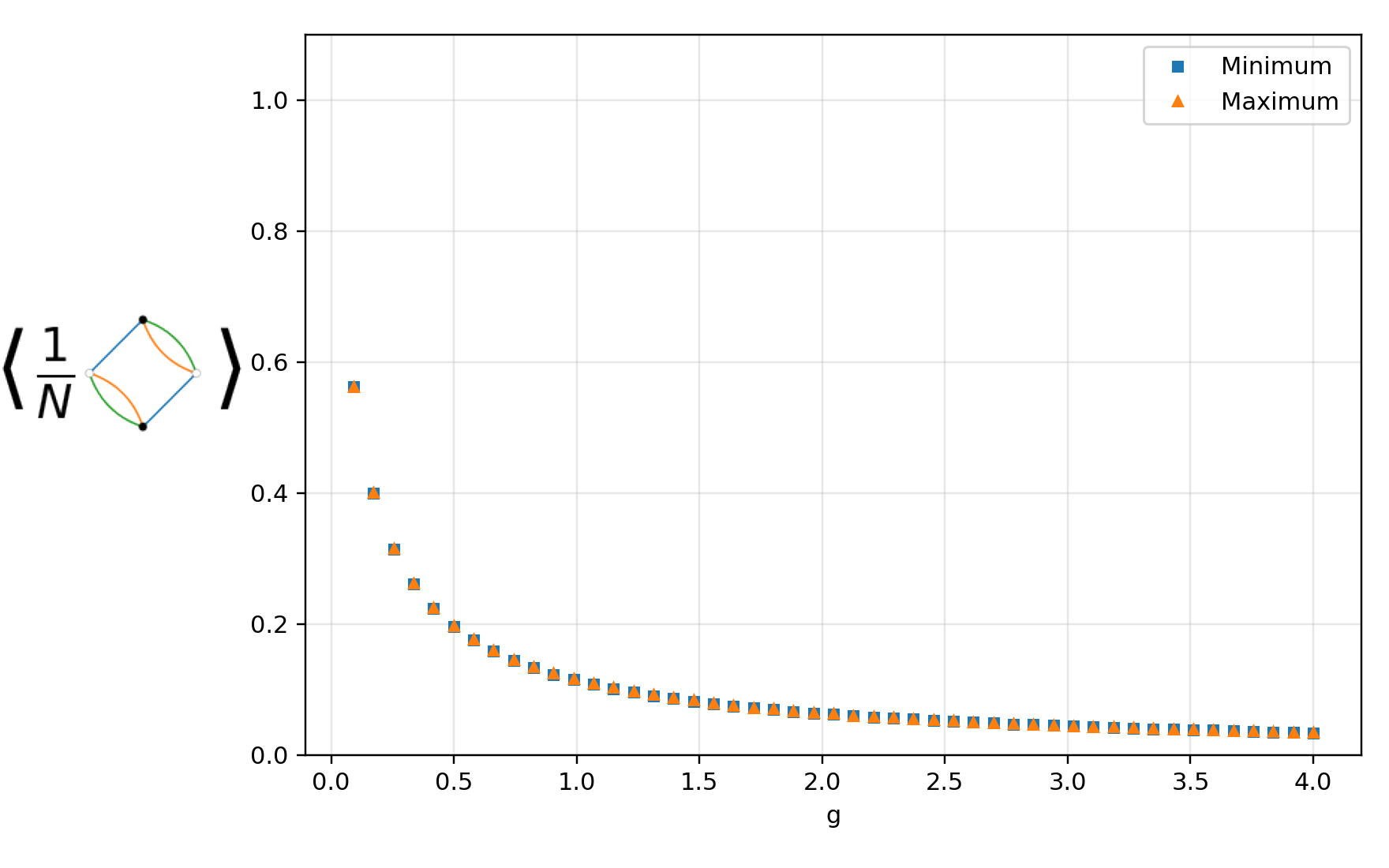}
    \end{minipage}

    \caption{
        Bounds on the $1/N$-rescaled expectation values of the elementary melon (top left) and the three quartic pillows (top right, bottom left, and bottom right) as a function of the quartic coupling $g$. For each plot, the bounds are obtained for $N = 5$.
    }
    \label{fig:three_pillows_results}
\end{figure*}

In Figure \ref{fig:three_pillows_results}, we showcase bounds found using this approach for the expectation value of the elementary melon and the three quartic pillows. By imposing that the Gram matrices are positive semi-definite, we obtain sharp bounds on these observables for positive values of the coupling $g$.

Using bounds found from the present method, we also compare how finite $N$ bounds compare with the large $N$ limit.

\begin{figure*}[t]
	\centering
	\begin{minipage}{0.3\textwidth}
		\centering
		\includegraphics[width=\linewidth]{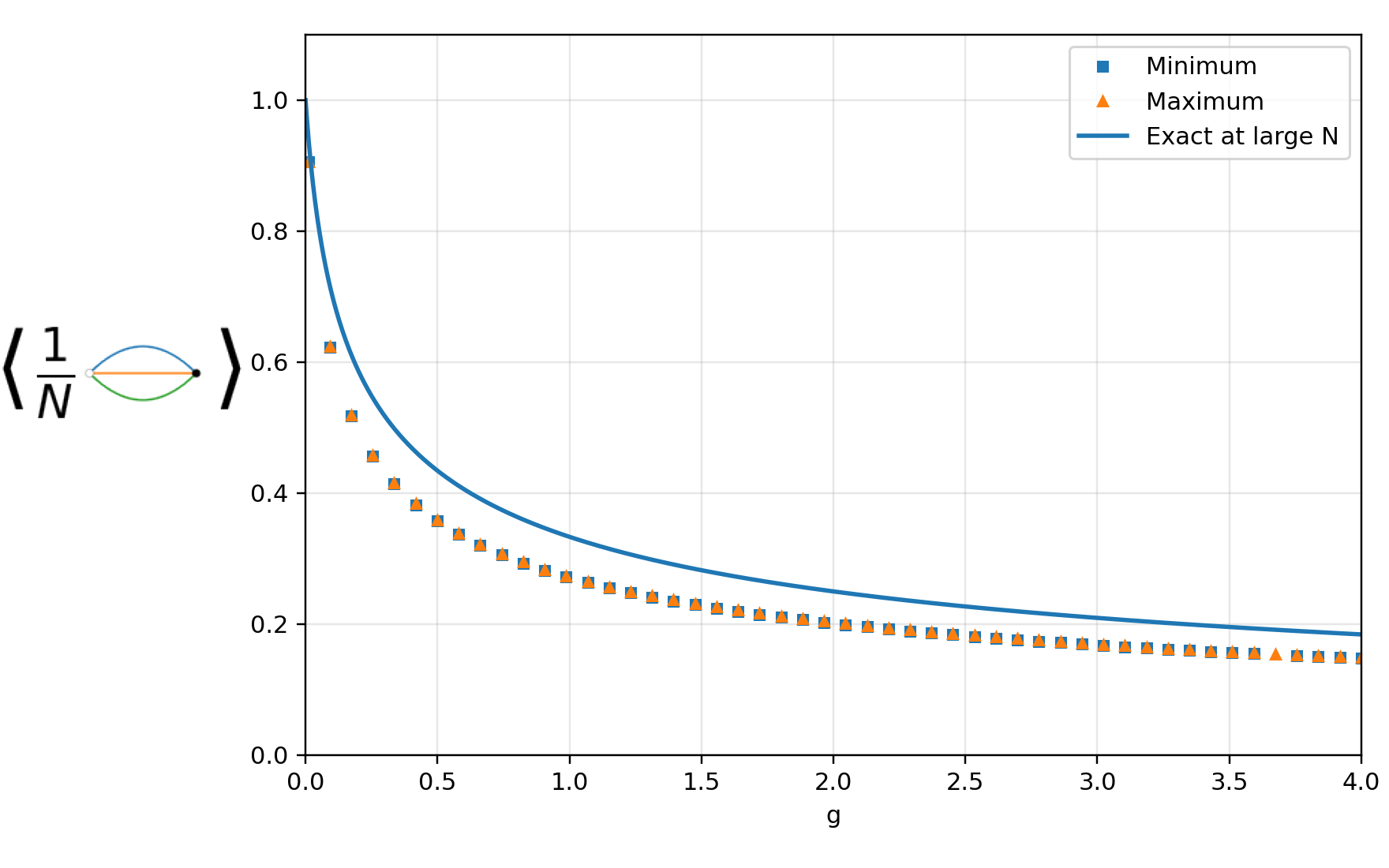}
	\end{minipage}
	\hfill
	\begin{minipage}{0.3\textwidth}
		\centering
		\includegraphics[width=\linewidth]{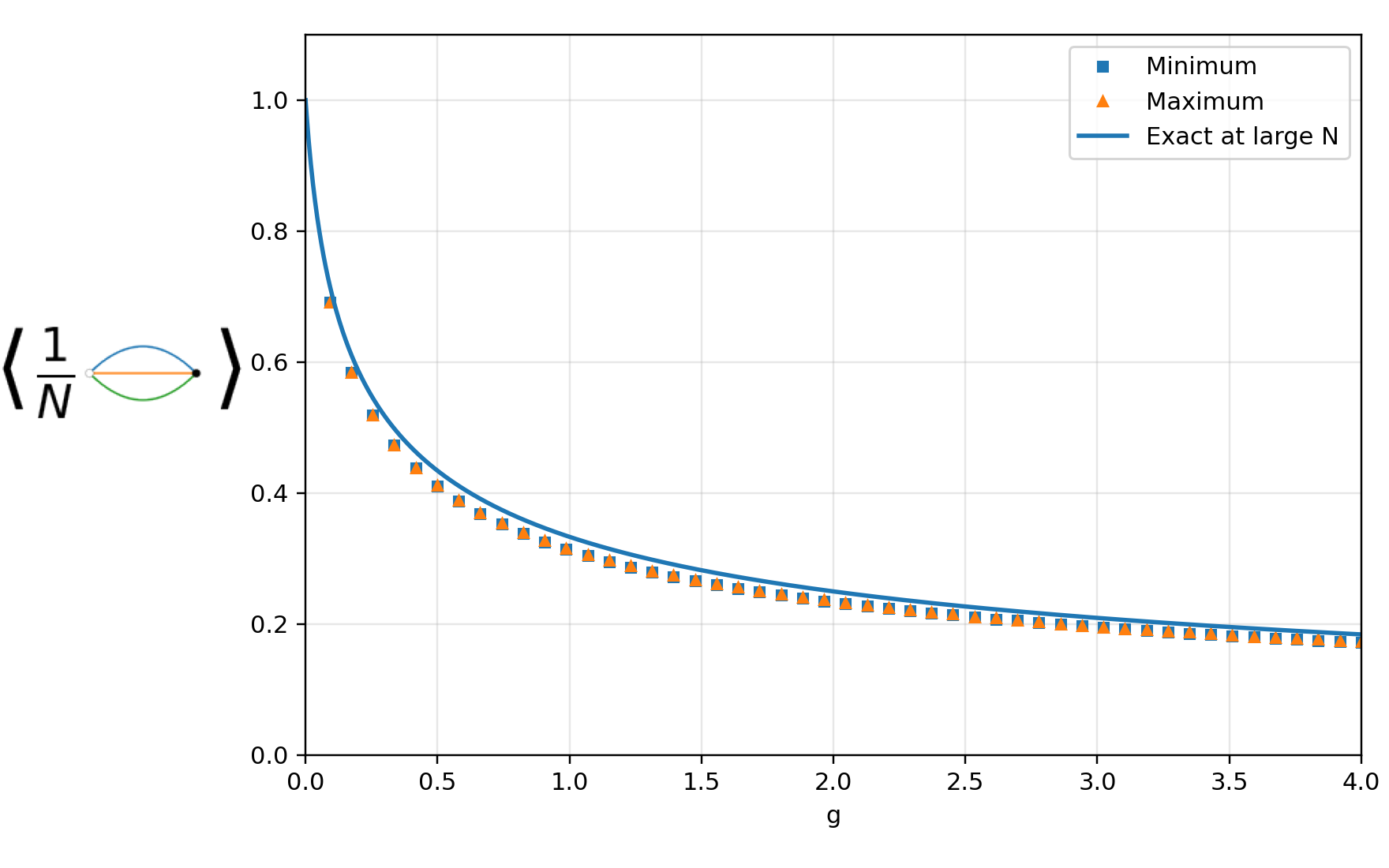}
	\end{minipage}
	\hfill
	\begin{minipage}{0.3\textwidth}
		\centering
		\includegraphics[width=\linewidth]{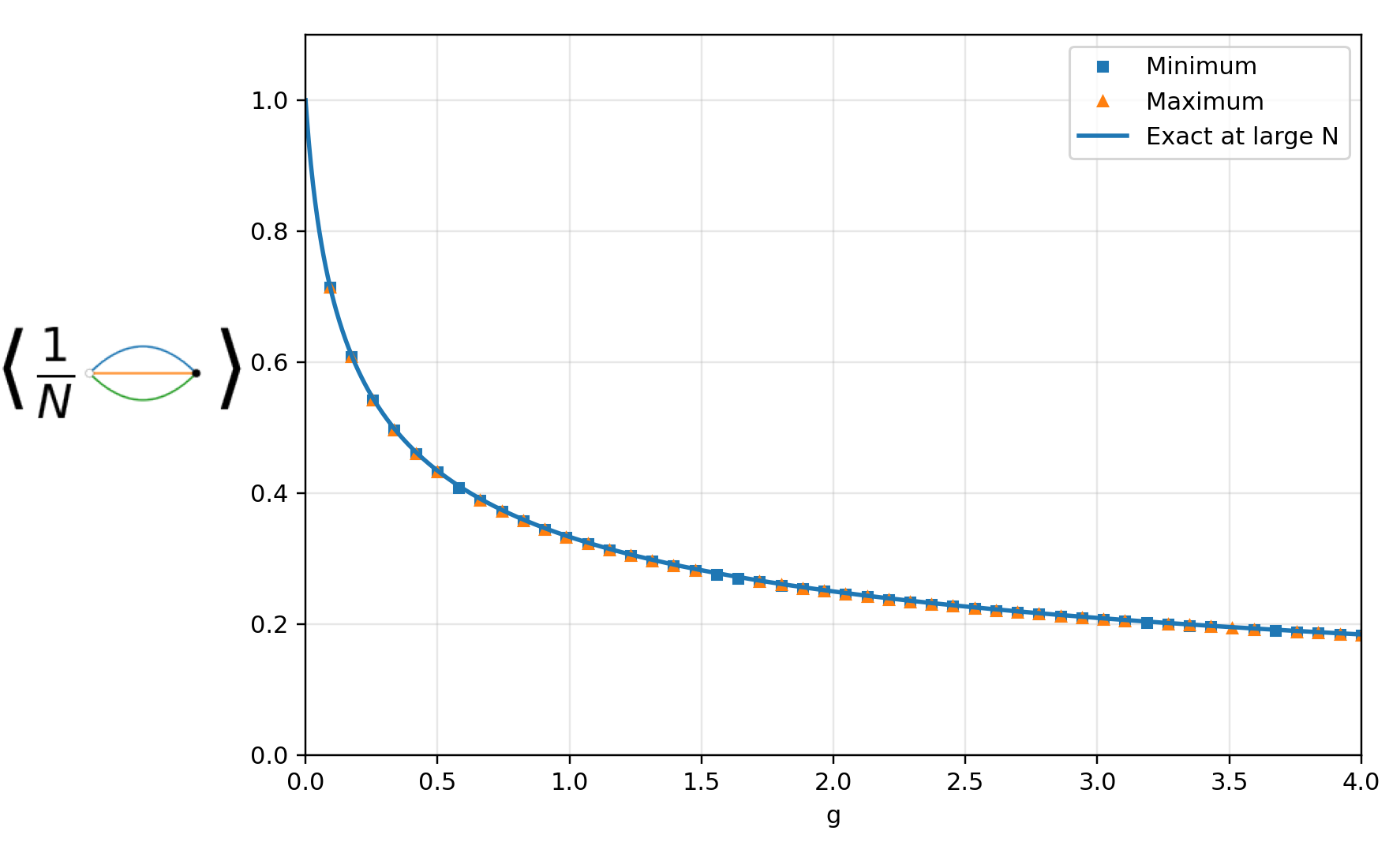}
	\end{minipage}
	\hfill
	\begin{minipage}{0.3\textwidth}
		\centering
		\includegraphics[width=\linewidth]{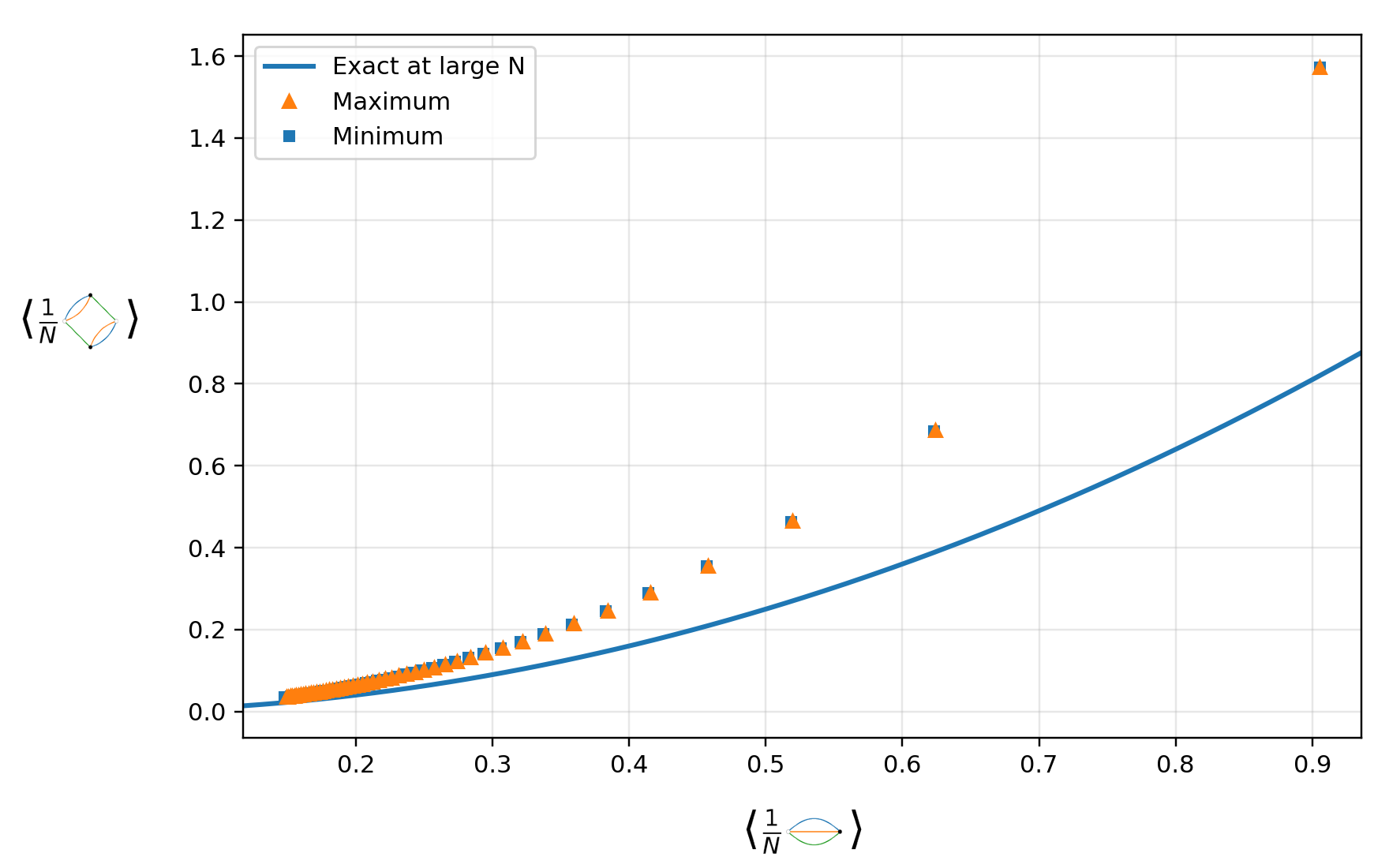}
	\end{minipage}
	\hfill
	\begin{minipage}{0.3\textwidth}
		\centering
		\includegraphics[width=\linewidth]{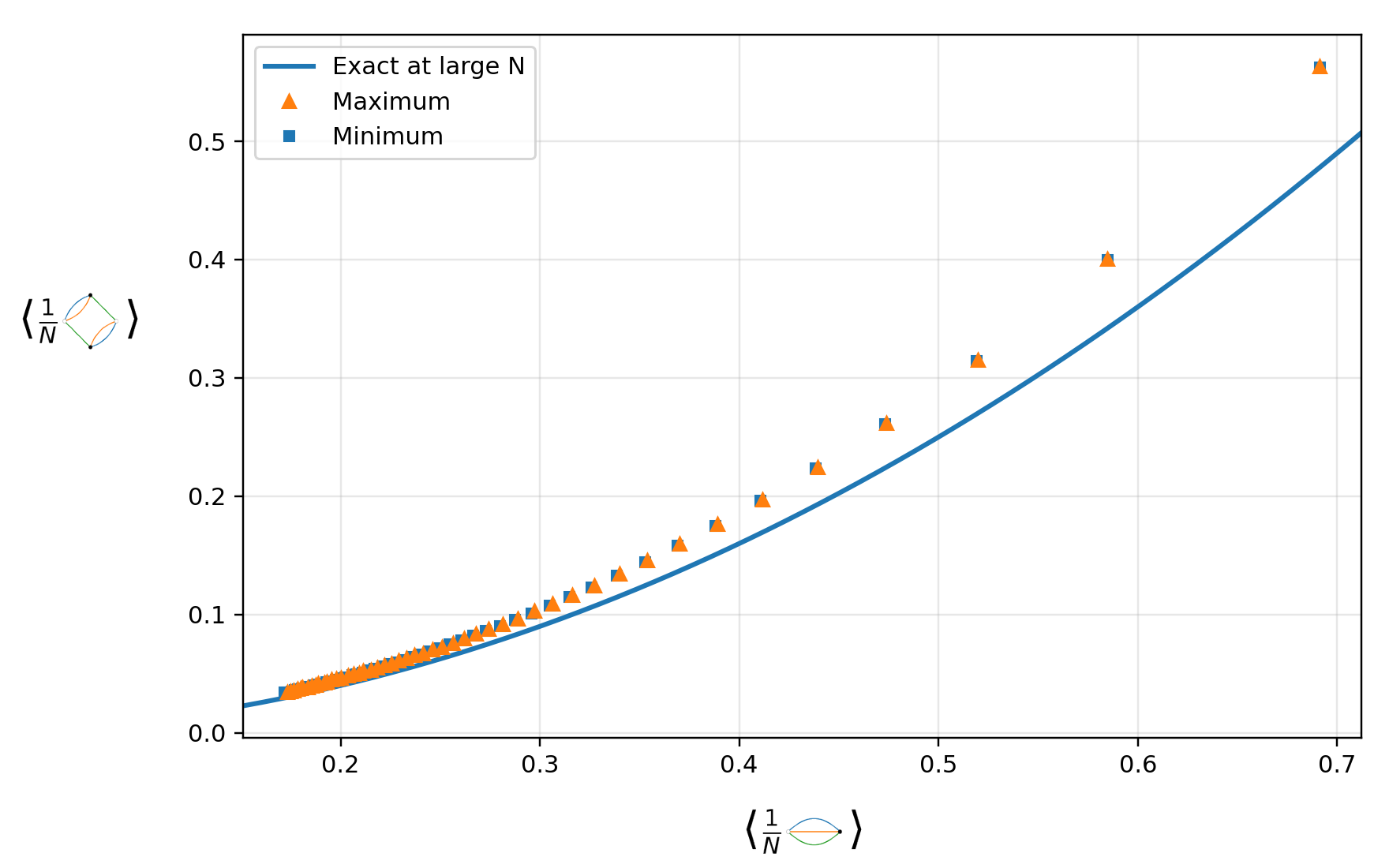}
	\end{minipage}
	\hfill
	\begin{minipage}{0.3\textwidth}
		\centering
		\includegraphics[width=\linewidth]{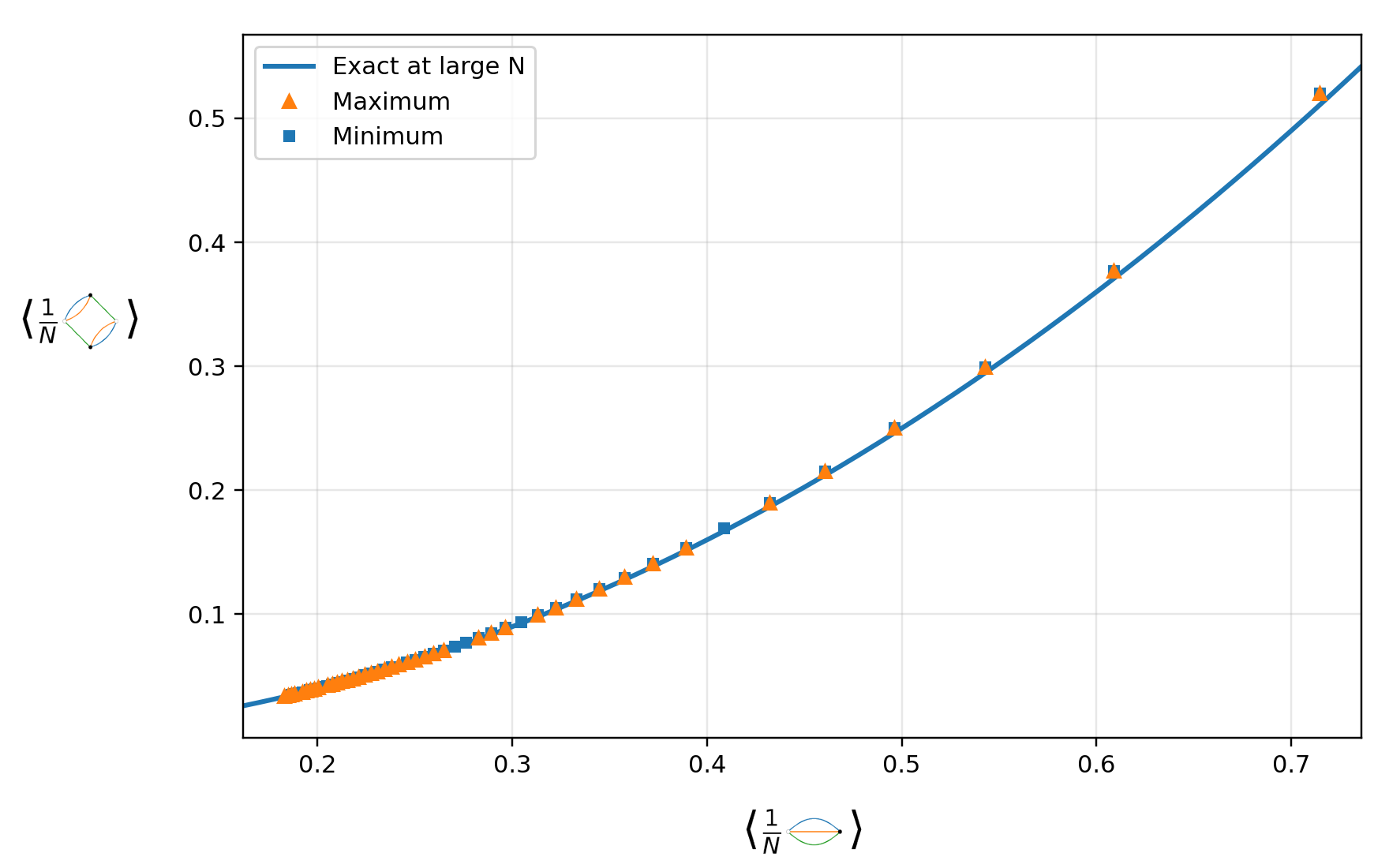}
	\end{minipage}
	
	\caption{
		Top row: Bounds on the $1/N$-rescaled expectation value of the elementary melon as a function of $g$. The bounds at $N = 1$ (top left), $N = 5$ (top center), and $N = 50$ (top right) are compared to the exact solution in the large $N$ limit (solid blue line). Bottom row: Bounds on the $1/N$-rescaled expectation value of the first pillow as a function of the $1/N$-rescaled expectation value of the elementary melon. The bounds are compared to the expected curve from Gaussian universality (solid blue line) for $N = 1$ (bottom left), $N = 5$ (bottom center), and $N = 50$ (bottom right).
	}
	\label{fig:large_N_tests}
\end{figure*}

In Figure \ref{fig:large_N_tests}, we compare the expectation value of the elementary melon to its exact solution in the large $N$. This exact solution was computed using Equation \ref{eq:melon_largeNexact} for $n = 3$. For $N = 1$ and $N = 5$, the bounds are found below the exact large $N$ result, and converge to this curve as one increases the value of $N$. For $N = 50$, the bounds become a close match with the exact large $N$ result.

In Figure \ref{fig:large_N_tests}, we also compare the expectation value of one of the pillows as a function of the elementary melon to the exact curve from Gaussian universality in the large $N$ limit. This is done by first computing the bounds for one of the pillows and the elementary melon for various values of $g$, then plotting the maximum and minimum allowed values for the pillow vs melon curve given these bounds. For $N = 1$ and $N = 5$, the bounds are found above the exact large $N$ curve found from Gaussian universality \eqref{eq:gauss_univ}, and converge to this curve as one increases $N$. For $N = 50$, the bounds become a close match with this curve.

For the three-pillow model, we thus find new and sharp numerical results at finite $N$, which converge to the appropriate curve in the large $N$ limit. We also find sharp bounds on deviations from Gaussian universality at finite $N$. In this case as well, the bounds converge to the exact curve in the large $N$ limit.

\noindent
\textbf{Example 2 - One pillow interaction:} Let us now demonstrate our approach with a model with only one quartic pillow as interaction. We will consider a model with the potential
\be
S = \melon + g \, \pillowone \, .
\label{eq:one_pillow_int}
\ee

This model is interesting because it breaks the symmetry between the three possible quartic pillows in the interaction term. To illustrate what we mean here, let us consider again the model with the potential (\ref{eq:three_pillow_int}). This potential is invariant under the transformations
\be
\pillowone \quad \leftrightarrow \quad \pillowtwo \quad \leftrightarrow \quad \pillowthree \, .
\ee
Therefore, we expect the expectation value of these three quantities to return the same value at any $g$ and any $N$\footnote{We checked, using results from Figure \ref{fig:three_pillows_results}, that this is the case.}. The potential (\ref{eq:one_pillow_int}) explicitly breaks this symmetry. Instead of having invariance under permutation of any pillows, we are left with invariance under permutation of the second and third pillow
\be
\pillowtwo \quad \leftrightarrow \quad \pillowthree \, .
\ee
This can be seen trivially from the fact that these two pillows simply do not appear in the tensor potential \eqref{eq:one_pillow_int}. From this permutation symmetry, we expect the expectation value of the pillows that are not present in the potential to return the same value at any $g$ and any $N$. However, the expectation value of the pillow present in the potential should return a different value at finite $N$ for $g \not= 0$.

One should note that this symmetry is only present at finite $N$. At large $N$, Gaussian universality enforces that all pillow expectation values must be the same according to  \eqref{eq:gauss_univ}, letting us recover permutation symmetry between all pillows. Therefore, observing this symmetry would showcase another form of deviation of Gaussian universality at finite $N$.

Given that this symmetry is a specific feature of the finite $N$ limit, it is natural to ask if the bootstrap method allows us to observe it. To probe this question, we found bounds on the expectation values of the three quartic pillows for the present system, applying the bootstrap method in the exact same way as for the three-pillow system \eqref{eq:three_pillow_int}. We generated 4 Gram matrices found by opening or cutting open bubbles with 6 vertices or, in other words, 3 black/white vertex pairs. This yielded the same $17 \times 17$ and $18 \times 18$ Gram matrices as for the three-pillow system. We then computed all Schwinger-Dyson equations relevant for the present system. These were found from the set of open bubbles found by opening invariants containing up to 10 vertices, or 5 black/white vertex pairs. We then partially solved the Schwinger-Dyson equations by relating bubbles with a larger number of vertices to bubbles with a smaller number of vertices, substituted the results in the Gram matrices, and computed the maximum and minimum allowed values of the expectation values of the three quartic pillows for various values of the coupling $g$, imposing that the Gram matrices must be positive semi-definite.

\begin{figure*}[t]
	\centering
	\begin{minipage}{0.3\textwidth}
		\centering
		\includegraphics[width=\linewidth]{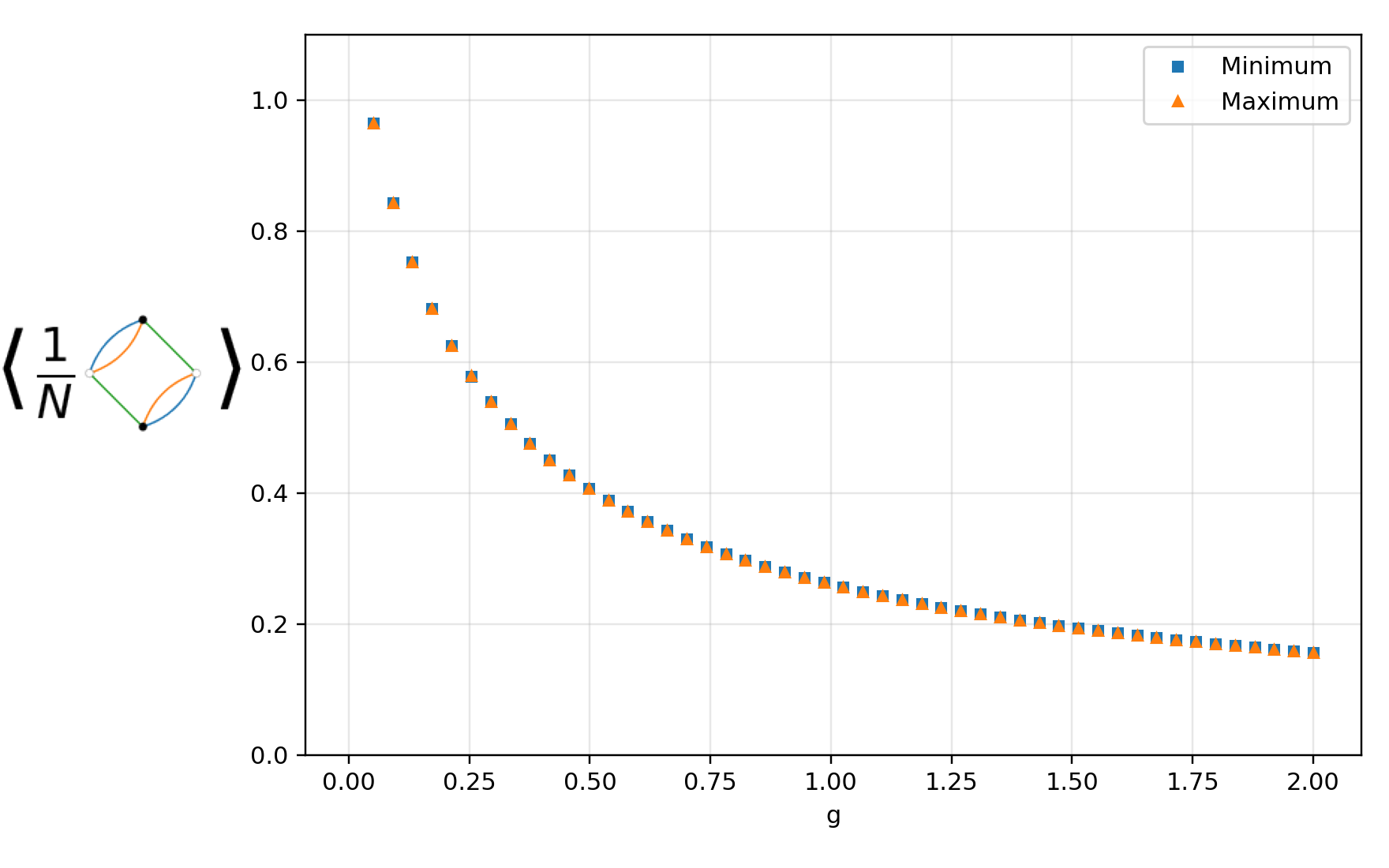}
	\end{minipage}
	\hfill
	\begin{minipage}{0.3\textwidth}
		\centering
		\includegraphics[width=\linewidth]{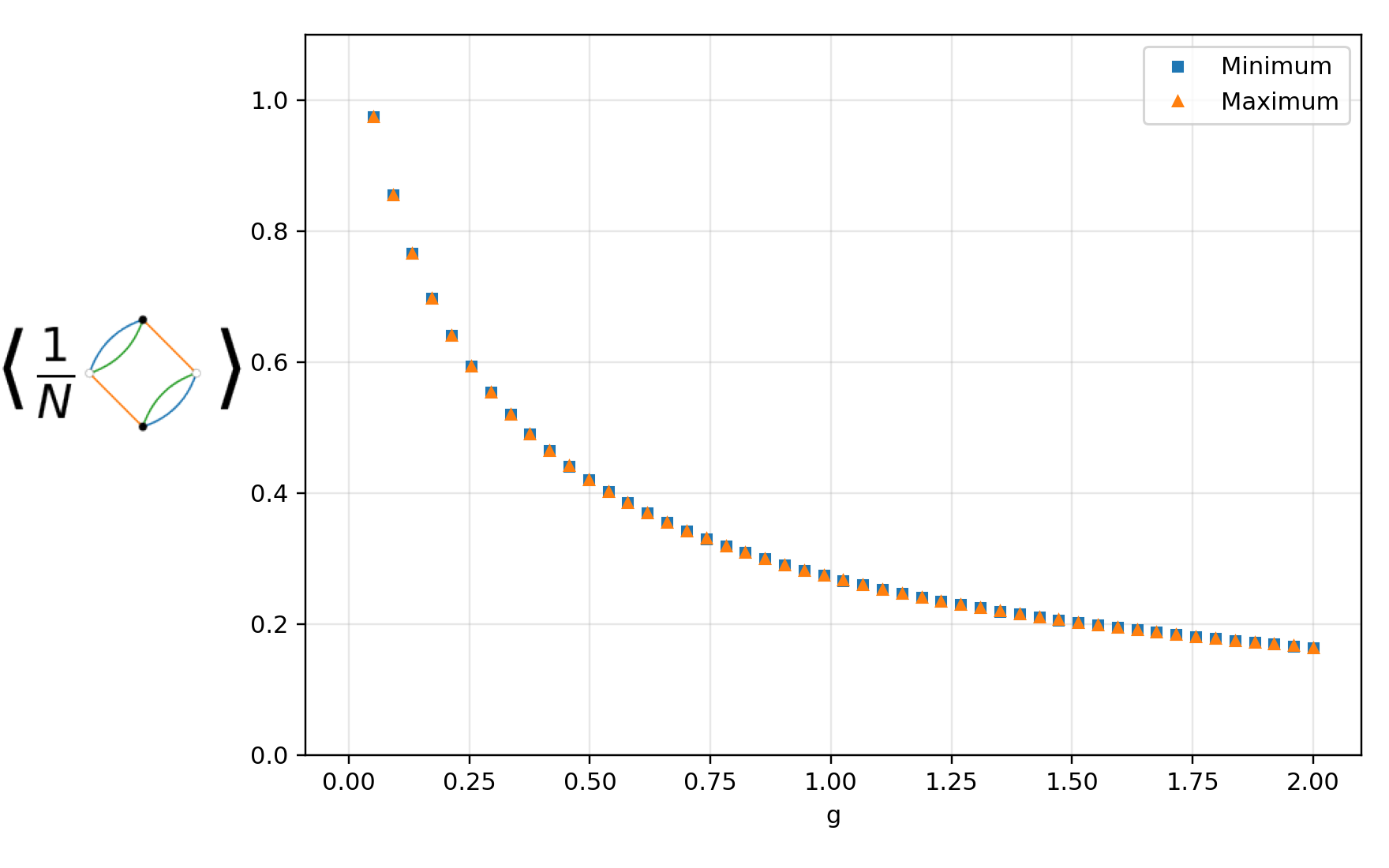}
	\end{minipage}
	\hfill
	\begin{minipage}{0.3\textwidth}
		\centering
		\includegraphics[width=\linewidth]{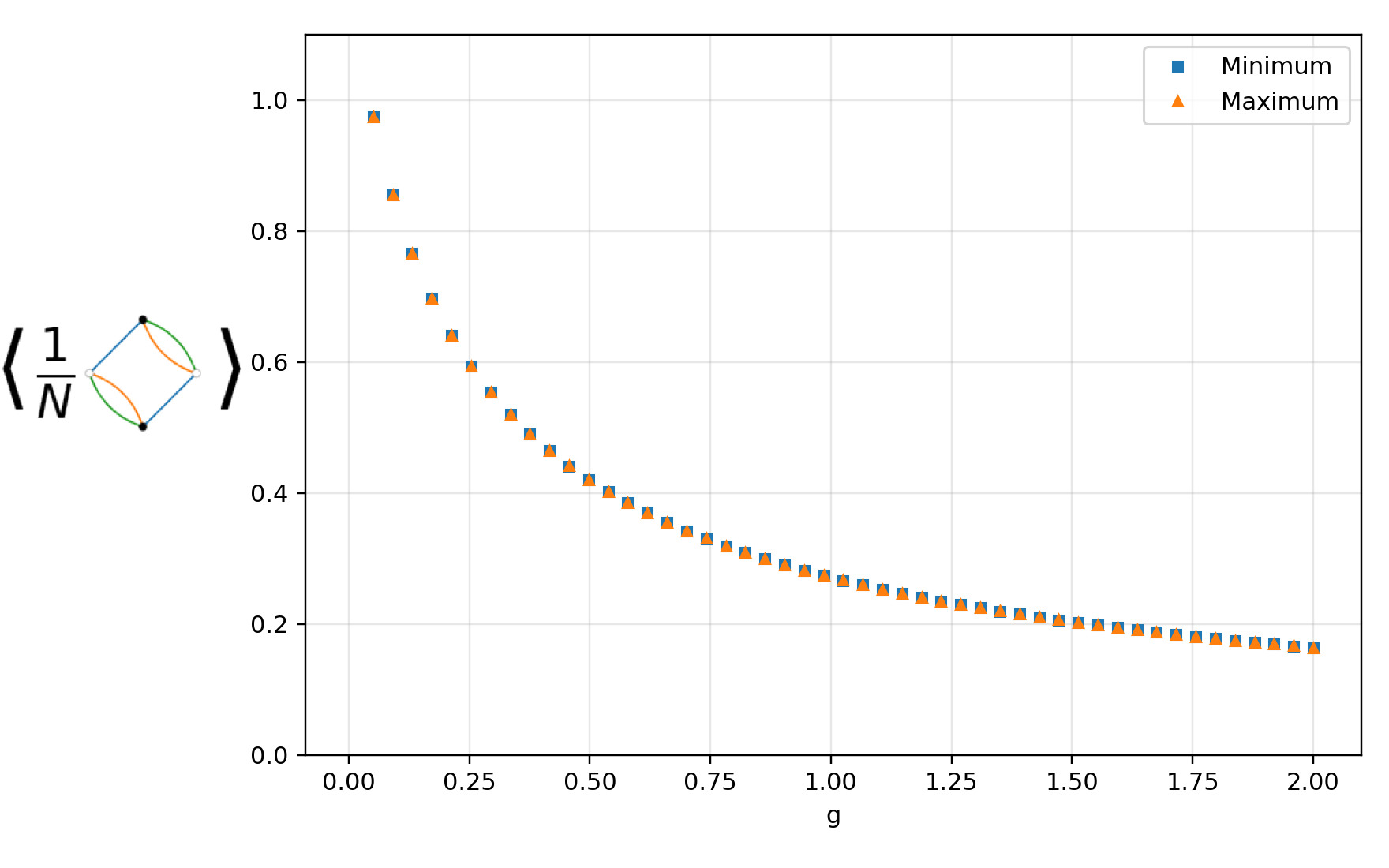}
	\end{minipage}
	\hfill
	\begin{minipage}{0.3\textwidth}
		\centering
		\includegraphics[width=\linewidth]{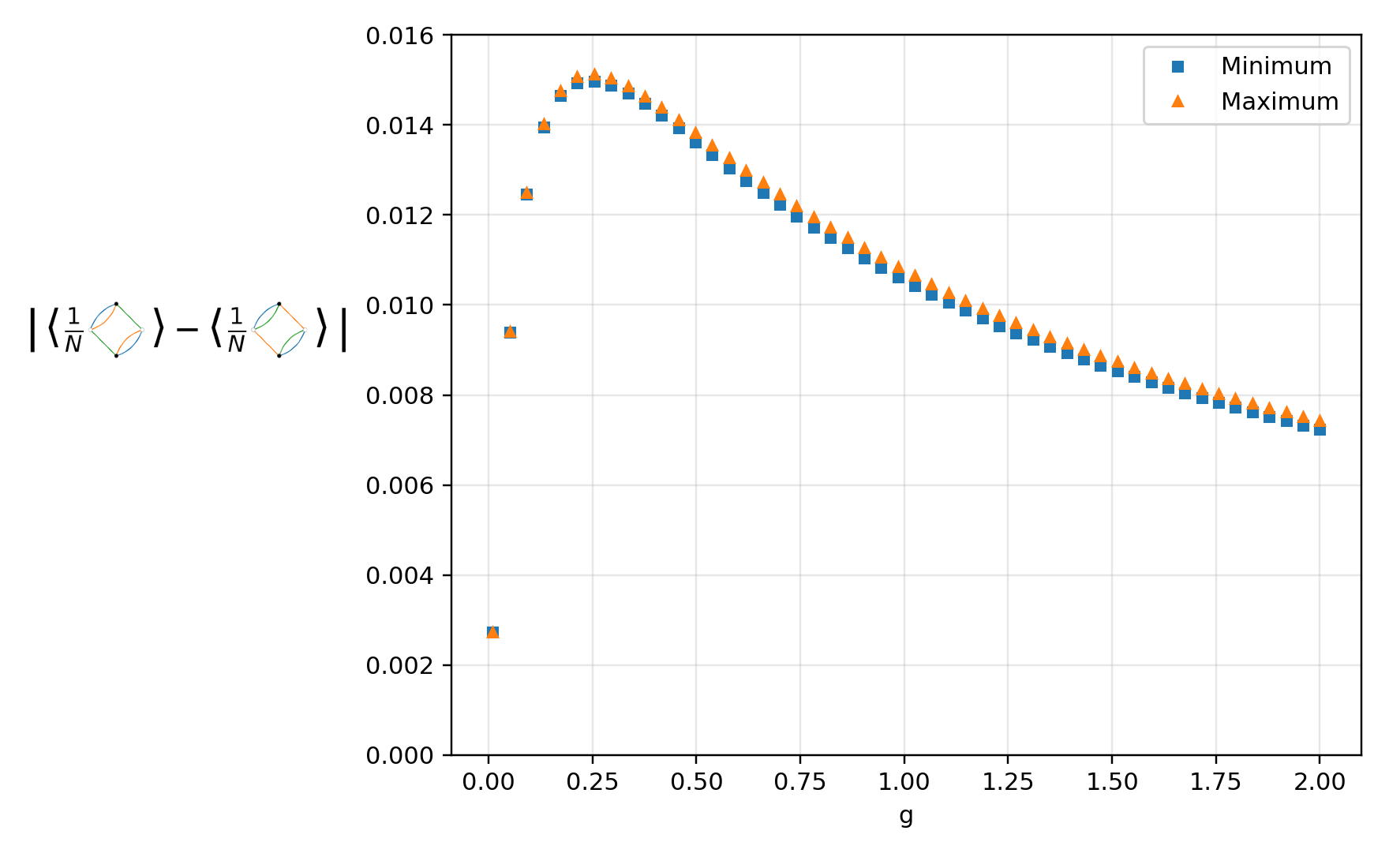}
	\end{minipage}
	\hfill
	\begin{minipage}{0.3\textwidth}
		\centering
		\includegraphics[width=\linewidth]{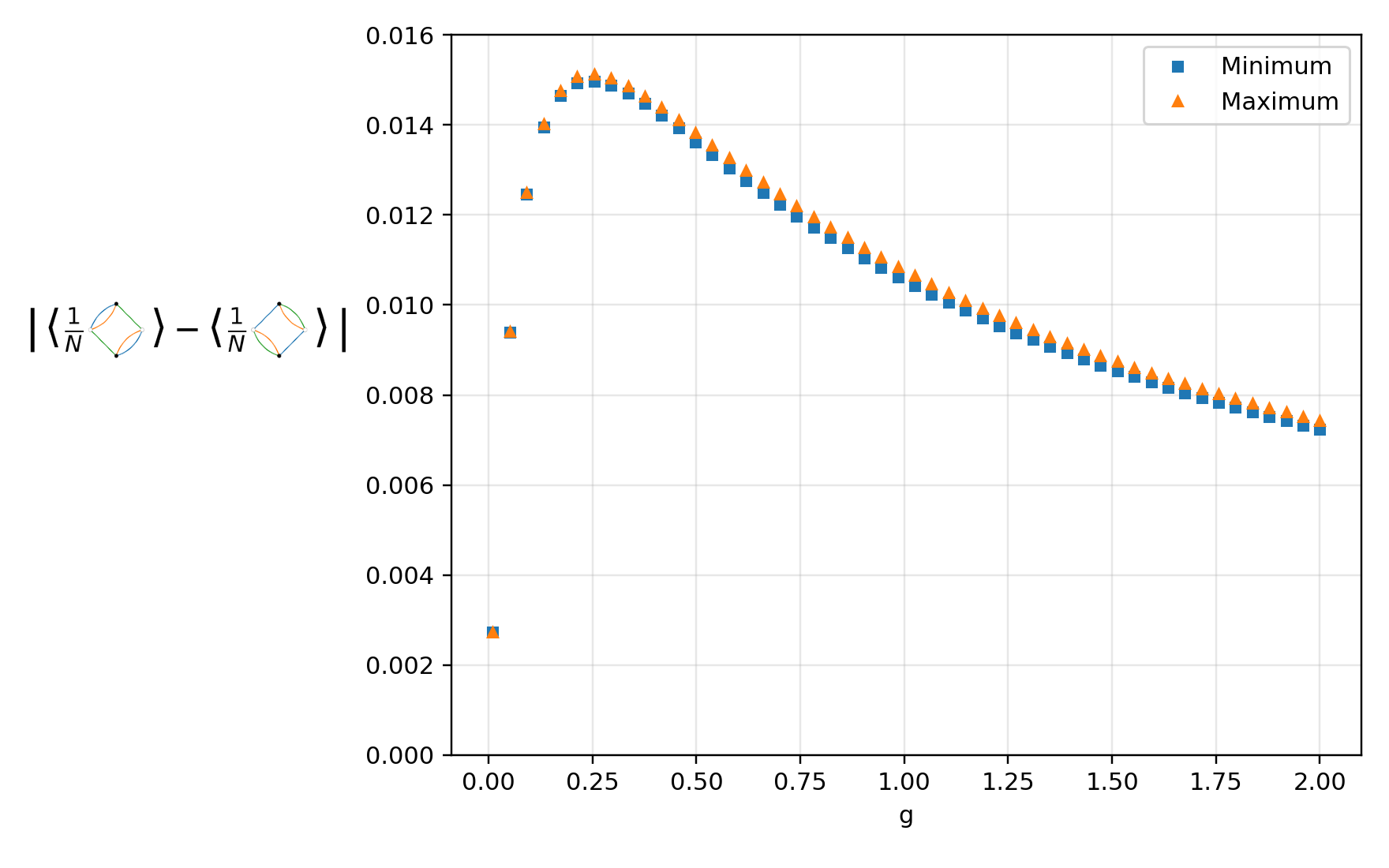}
	\end{minipage}
	\hfill
	\begin{minipage}{0.3\textwidth}
		\centering
		\includegraphics[width=\linewidth]{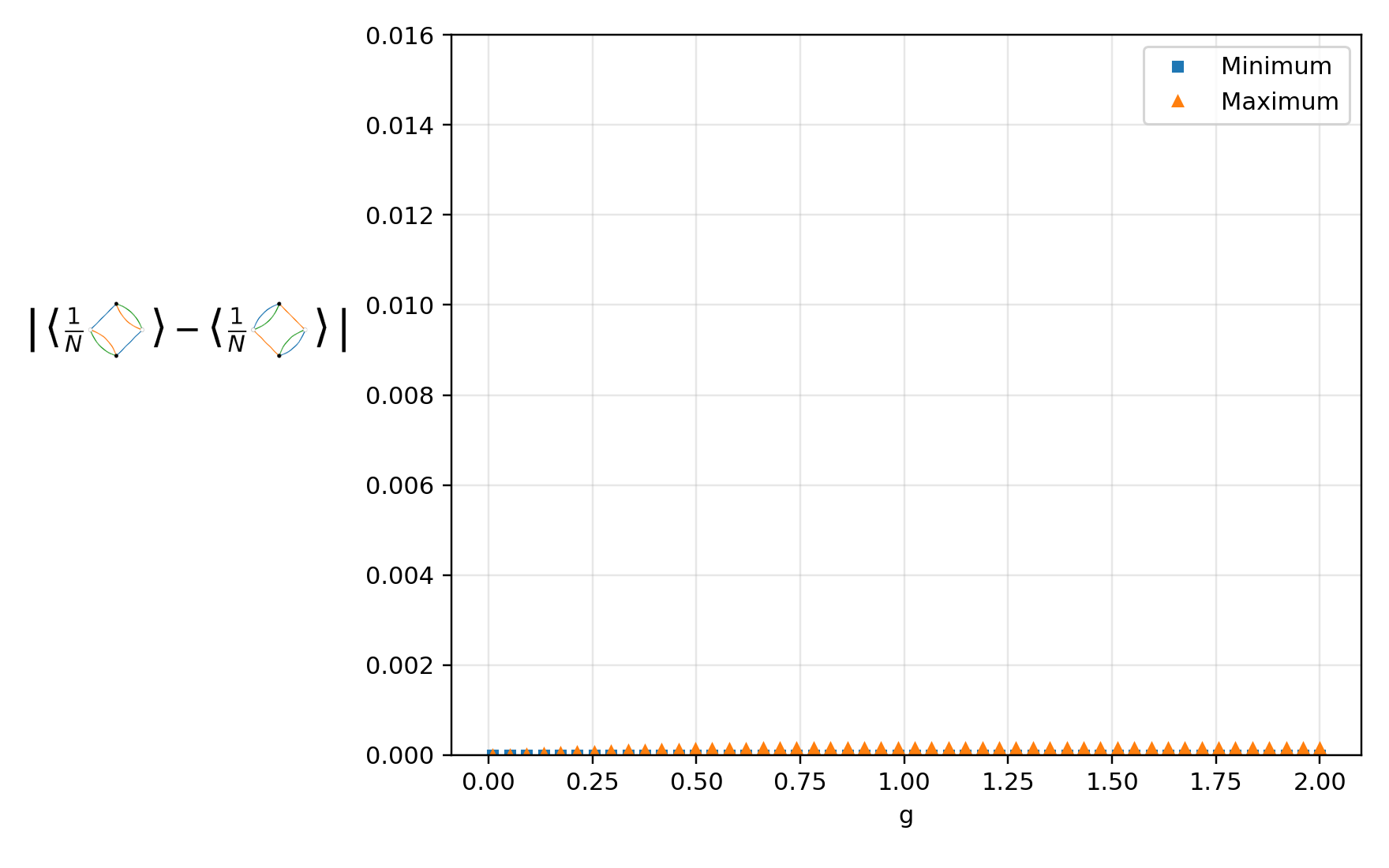}
	\end{minipage}
	
	\caption{
		Top row: Bounds on the $1/N$-rescaled expectation values of the three quartic pillow invariants for $N=5$. Bottom row: Bounds on the $1/N$-rescaled differences between the expectation values of these quartic pillow invariants for $N=5$.
	}
	\label{fig:one_pillows_results}
\end{figure*}

In Figure \ref{fig:one_pillows_results}, we showcase the results found by computing the expectation value of all three pillows at $N = 5$ for varying values of $g$. For all three pillows, we find sharp bounds on the expectation value as a function of the coupling $g$, which gives very similar results. However, by computing the difference between the expectation values of the pillows, one can see deviations reflecting the symmetry of the system. For example, if we compute the bounds on the difference between the first pillow and the second pillow and the bounds on the difference between the first and the third pillow, one observes small deviations. For the bounds on the difference between the last two pillows, however, no deviation is observed as expected from the permutation symmetry of the present system.

We thus conclude that the bootstrap lets us recover sharp bounds on the expectation values of the one-pillow system. Moreover, we recover another signature of the breakdown of Gaussian universality, namely the re-emergence of pillow permutation symmetry at finite $N$.

\section{Conclusion and Discussion}
\label{sec:conclusion}

We explored new positivity constraints for tensor models constructed from "open bubbles" and "color matrices". Using these constraints, we found sharp constraints at finite $N$, and probed deviations from Gaussian universality in this limit.

To test our approach, we studied a rank 3 Gaussian model with a three pillow quartic interaction, and a rank 3 Gaussian model with one quartic pillow as interaction. 

For the three-pillow model, we found sharp bounds on the expectation values of the elementary melon and the three quartic pillows for positive coupling at finite $N$. Using these bounds, we studied finite $N$ deviations from the large $N$ limit for the elementary melon expectation value. We found that, for increasing values of $N$, the large $N$ behavior is recovered. We also studied the deviation from Gaussian universality at finite $N$. When plotting the bounds of one of the quartic pillows vs. the elementary melon, we found that the finite $N$ curve lies above the expectation from Gaussian universality at large $N$. As one increases $N$, the curve converges to the expected large $N$ behavior.

For the model with a one-pillow quartic interaction, we also found sharp bounds on the expectation value of the three pillows at finite $N$. We then computed bounds for the difference between the three pillows, and found results matching the expected symmetries between the pillows at finite $N$. At large $N$, one expects no difference between the pillows as a consequence of Gaussian universality. Therefore, this result showcases additional expected deviations from Gaussian universality at finite $N$.

To obtain the present bounds, we will stress that we have not used any factorization properties in the large $N$ limit. The large $N$ limit is naturally recovered by using the Schwinger-Dyson equations at finite $N$ and the additional constraint of positivity, and taking the value of $N$ to be sufficiently large. For the models we studied, $N = 50$ seemed sufficient to obtain large $N$ behavior. However, this feature should not be taken as universal. We expect the large $N$ limit to be recovered at different values of $N$ depending on the model.

In the present work, we displayed results for rank 3 models with quartic interactions. The reason for this choice is that positivity bounds seem to yield stronger bounds for these classes of models with Gram matrices of smaller sizes. For models with higher order interactions and higher rank, one must consider larger Gram matrices to obtain the same level of accuracy. We also restricted our analysis to melonic interactions and expectation values of melonic graphs. Extending the study to non-melonic observables remains an important direction for future work and is necessary to assess the broader effectiveness of our approach. In this context, additional or qualitatively different constraints may be required to obtain comparably strong bounds.

A challenge of the tensor bootstrap is that the size of the Gram matrices, the number of Schwinger-Dyson equations that must be considered, and the runtime of the semi-definite programming algorithm increase combinatorially as one goes to higher constraint truncation level and higher tensor rank. Therefore, finding the minimal set of constraints that yield strong bounds is of paramount importance. In the present analysis, we found that level 3 constraints are already strong enough to yield sharp bounds. In this case, generating 4 plots with 50 points took about 30 minutes on a laptop, which we found reasonable. We also tried obtaining bounds for rank 4 tensors. In this case, finding one point for a level three constraint took about 7 hours of runtime on a laptop. To carry out our analysis on higher-order tensors, a more optimized approach might be needed. 

To decrease runtime, it would be interesting to investigate if bounds from a subclass of open bubbles and color matrices dominate the constraints. If so, this could help decrease the size of the Gram matrices, and therefore improve runtime. It would also be interesting to study the interplay of the bounds studied here and the ones studied in \cite{Pagliaroli:2026hxv}. 

On a tangent related to factorization properties of bubbles, such as Gaussian universality, it has recently been demonstrated that the expectation value of multi-trace observables does not always factorize in tensor models \cite{Gurau:2025evo,Berthold:2026zxk,Carrozza:2026cci}. In light of our results, where we demonstrate the non-factorizability of melonic bubbles at finite $N$, the bootstrap appears to be a tool of choice to probe the properties of non-factorizing observables. It would thus be interesting to use the present tools, or those presented in \cite{Pagliaroli:2026hxv}, to probe this matter in detail. To achieve this task, a better understanding of how the bootstrap approach applies to real tensors with a large number of vertices, and at large $N$, may be required. We leave these potential studies for future work.

\section{Acknowledgements}

S.L. would like to thank Nathan Pagliaroli for useful discussions.

\bibliography{cite}

\end{document}